\documentclass[aps,pra,superscriptaddress,twocolumn]{revtex4}

\usepackage[utf8]{inputenc}
\usepackage{amssymb}
\usepackage{amsmath}
\usepackage{bm}
\usepackage{mathtools}
\usepackage{bbold}
\usepackage{upgreek}
\usepackage{xfrac}
\usepackage{soul}
\usepackage{bm}
\usepackage{comment}
\usepackage{xcolor}
\usepackage{amsmath,amsfonts,amssymb}
\usepackage{mathtools}
\DeclarePairedDelimiter{\ceil}{\lceil}{\rceil}
\DeclarePairedDelimiter{\floor}{\lfloor}{\rfloor}
\usepackage{txfonts}
\DeclareSymbolFont{matha}{OML}{txmi}{m}{it}% txfonts
\DeclareMathSymbol{\varv}{\mathord}{matha}{118}
\usepackage{bbm}
\begin{document}

\title{Solving nonlinear differential equations with differentiable quantum circuits}

\author{Oleksandr Kyriienko}
\affiliation{Department of Physics and Astronomy, University of Exeter, Stocker Road, Exeter EX4 4QL, UK}
\affiliation{Qu \& Co B.V., PO Box 75872, 1070 AW, Amsterdam, The Netherlands}

\author{Annie E. Paine}
\affiliation{Department of Physics and Astronomy, University of Exeter, Stocker Road, Exeter EX4 4QL, UK}
\affiliation{Qu \& Co B.V., PO Box 75872, 1070 AW, Amsterdam, The Netherlands}

\author{Vincent E. Elfving}
\affiliation{Qu \& Co B.V., PO Box 75872, 1070 AW, Amsterdam, The Netherlands}

\date{\today}

\begin{abstract}
We propose a quantum algorithm to solve systems of nonlinear differential equations. Using a quantum feature map encoding, we define functions as expectation values of parametrized quantum circuits. We use automatic differentiation to represent function derivatives in an analytical form as differentiable quantum circuits (DQCs), thus avoiding inaccurate finite difference procedures for calculating gradients. We describe a hybrid quantum-classical workflow where DQCs are trained to satisfy differential equations and specified boundary conditions. As a particular example setting, we show how this approach can implement a spectral method for solving differential equations in a high-dimensional feature space. From a technical perspective, we design a Chebyshev quantum feature map that offers a powerful basis set of fitting polynomials and possesses rich expressivity. We simulate the algorithm to solve an instance of Navier-Stokes equations, and compute density, temperature and velocity profiles for the fluid flow in a convergent-divergent nozzle.
\end{abstract}

\maketitle

\section{Introduction}

Differential equations are ubiquitous in various fields of science~\cite{SimmonsBook}. Their applications range from predicting motion in mechanics and fluid dynamics~\cite{AndersonBook}, to describing reactions in chemistry, ecosystem balance in ecology, market dynamics in finance, and disease spreading in epidemiology. In many cases finding solutions to nonlinear systems of differential equations (DEs) is challenging, and requires advanced numerical techniques~\cite{ButcherBook,BoydBook}. The difficulties arise in DEs for many functions and variables (known as `the curse of dimensionality'), models with high degree of nonlinearity, optimal control problems, as well as stiff and numerically unstable systems \cite{HairerBook,GearRev}. Classical numerical methods for solving DEs can be divided into local and global methods. Local approaches rely on discretization of the space of variables, with derivatives being approximated with numerical differentiation techniques (finite differencing and Runge-Kutta methods). Often a fine grid for multivariable functions is required to represent a solution qualitatively and quantitatively~\cite{ButcherBook}, leading to increasing computational cost. Global methods represent the solution in terms of a suitable basis set~\cite{BoydBook}. This recasts the problem to finding optimal coefficients for the polynomial approximation (e.g. Fourier or Chebyshev) of the sought function. Finding spectral solutions for complex problems may require ever-increasing basis sets to achieve high accuracy, enlarging the global differential operator, and introduces difficulties when dealing deterministically with boundary conditions.

Quantum computers offer a fundamentally different approach to perform computation, and possess algorithmically superior scaling for certain problems that include amplitude amplification and Abelian hidden subgroup problems~\cite{NielsenChuang,MontanaroRev}. For prototypical linear algebra tasks, quantum computers can be used as exponential accelerators for solving linear system of equations (LSE), as offered by the HHL algorithm \cite{HHL2009} and other quantum LSE approaches with improved scaling~\cite{Childs2017, Wossnig2018, Subasi2019, LinLin2019, Woerner2020}. These developments, among others, have led to the nascent field of quantum machine learning~\cite{Biamonte2017, SchuldRev, Lloyd2014, Rebenfrost2014, Lloyd2016, Dunjko2016, Schuld2016, Nana2018, Rebentrost2019, Blank2020}. Quantum LSE solvers can be applied to linear differential equations that are rewritten as systems of algebraic equations using a finite differencing scheme (Euler's method). Several studies have developed generic quantum solvers for linear differential equations, and have shown algorithmic improvements in terms of accessed quantum oracles~\cite{Leyton2008, Cao2012, Berry2014, Montanaro2016, Berry2017, Costa2019, Childs2020b, Wang2020, Linden2020}. For the global numerical differentiation methods a spectral solver was proposed in \cite{Childs2020a}. The power of HHL-based algorithms relies on \emph{amplitude encoding}, where a function is represented as a quantum state $|u\rangle = \sum_{k=1}^{2^N} u_k |k\rangle$ encoded in complex $u_k$ amplitudes of computational basis states $|k\rangle$ for $N$ qubits. This allows for compressing large grids into a small qubit register, providing exponential memory advantage. However, there are several caveats that should be considered \cite{HHL2009}. First, the proposed quantum algorithms solve an equivalent quantum task of finding a state $|u\rangle = \hat{A}^{-1} |b\rangle$ given the matrix $\hat{A}$ acting on pre-defined (boundary) state $|b\rangle$. Reading out the function values from the quantum state $|u\rangle$ in general requires exponential sampling, representing a so-called data `output problem'. Second, the preparation of a generic input state $|b\rangle$ requires sophisticated techniques such as quantum random access memory (QRAM) \cite{Giovannetti2008}, thus leading to a so-called data `input problem'. Finally, translating quantum oracles that encode DEs (matrix $\hat{A}$) can introduce huge computational overheads, making simulation infeasible even for a fault-tolerant quantum computer~\cite{Scherer2017}.

Modern quantum processors are not perfect, and hardware available today correspond to noisy intermediate scale quantum (NISQ) devices~\cite{Preskill2018}. They operate for a limited circuit depth, but still enjoy a Hilbert space that increases exponentially with the number of qubits. Recently, such a setup was shown to provide advantage over classical methods for a tailored problem~\cite{Arute2019}. However, finding NISQ protocols that can offer advantage for industrially relevant tasks represents an open challenge. One promising direction corresponds to simulation of material science and chemistry, where hybrid quantum-classical approaches are developed for preparing ground states of correlated molecules~\cite{OxfordRev, ZapataRev, Elfving2020b}. The paradigmatic hybrid algorithm corresponds to the variational quantum eigensolver (VQE) \cite{Peruzzo2014, McClean2016}, where a quantum computer is used to prepare parametrized quantum states and measure their energy, while classical optimization is used for finding the best variational parameters. This strategy enables quantum calculations with relatively noisy devices, and allows for numerous advances from the experimental~\cite{OMalley2016, Kandala2017, Colless2018, Kokail2019, Ganzhorn2019, McCaskey2019, IonQ2019, YamamotoExp, Arute2020} and theoretical~\cite{Gard2019, Grimsley2019, Tang2019, Claudino2020, Huggins2019b, Tzu-Ching2020, Kubler2020, Nakanishi2019, Metcalf2020, Bauman2020, OBrien2019, Parrish2019a, Yalouz2020, Elfving2020a} perspectives.

The rise of VQE has launched further development of variational quantum algorithms (VQAs)~\cite{BenedettiRev, AlejandroRev, Huggins2019, Cerezo2020, Lamata2020}, targeting generic data analysis and machine learning tasks. Performed by \emph{parametrized quantum circuits}, also dubbed as quantum neural networks (QNNs) \cite{BenedettiRev}, VQAs rapidly expand applications to include classification \cite{Farhi2018, Mitarai2019, Havlicek2019, SchuldBocharov2020, Suzuki2020, Yano2020, DuNana2020}, regression \cite{Mitarai2019}, generative modelling \cite{JGLiu2018, LloydWeedbrook2018, Zeng2019, Verdon2017, BenedettiGrant2019, BenedettiAlejandro2019, Zoufal2019, Romero2019, Coyle2020}, clustering \cite{rigetti2017, Mendelson2019}, reinforcement learning~\cite{DunjkoRev, Melnikov2018, Lamata2018, Melnikov2020, Wallnofer2020, OS2020}, quantum simulation \cite{Kreula2016, Cirstoiu2020, Endo2020, Paulson2020},  quantum metrology \cite{Meyer2020, Ouyang2019, Koczor2020} and other tasks. The power of VQAs for machine learning (ML) tasks comes from the high expressibility of quantum circuits, where data points are mapped to quantum states. Also, a large role for designing efficient algorithms is played by automatic differentiation and natural gradient techniques~\cite{Mitarai2019, Schuld2019, Mitarai2020, Parrish2019b, McArdle2018, Stokes2020, Yamamoto2019, Koczor2019, Sweke2020, Koczor2020b}, developing deep networks and new QNN architectures \cite{Verdon2018,Beer2020,Garg2020,Cong2019}, as well as proposing operationally meaningful cost functions for optimization~\cite{Cerezo2020,Cerezo2020b}, and optimal ansatz search \cite{Herasymenko2019, Skolik2020, Sim2020, Du2020, Tsinghua2020, ZhangHsieh2020}. First experiments show promising results for systems of small and increasing size \cite{Havlicek2019, Zhu2019, Huang2020}. Various strategies of error mitigation were proposed that can further improve the performance of algorithms when run on physical devices \cite{Li2017, Temme2017, Endo2018, Bonet-Monroig2018, McArdle2019, Otten2019, Bondarenko2020, Giurgica-Tiron2020}. Finally, considering linear algebra problems, several VQAs were also proposed to solve LSEs \cite{Huang2019, Bravo-Prieto2020, Xu2020, Chen2019, Perelshtein2020, borle2020}, and ongoing efforts are directed towards improving their workflow. To date, several small scale linear equation solvers have been implemented using a nuclear magnetic resonance-based setup \cite{Pan2014,Xin2020}.

As for \emph{nonlinear} differential equation solvers, available quantum protocols remain scarce and the field is only starting to grow. An intriguing approach was proposed in Ref.~\cite{Lubasch2020}, where nonlinearity is introduced using quantum nonlinear processing units realized by ancillary quantum registers and controlled-multiqubit operations. The approach uses amplitude encoding, offering memory saving while potentially facing the data input and output problems. Another hybrid quantum-classical approach is presented in Ref.~\cite{Gaitan2020}, where computational fluid dynamics applications are considered. This relies heavily on  classical computation, with the quantum processor being delegated to perform function averaging (using amplitude estimation as subroutine). Using an oracle-based approach, amplitude-encoded states, and two quantum Fourier transforms, the discussed algorithm remains an option more suitable for future fault-tolerant devices.
%%%
\begin{figure*}
\includegraphics[width=0.8\linewidth]{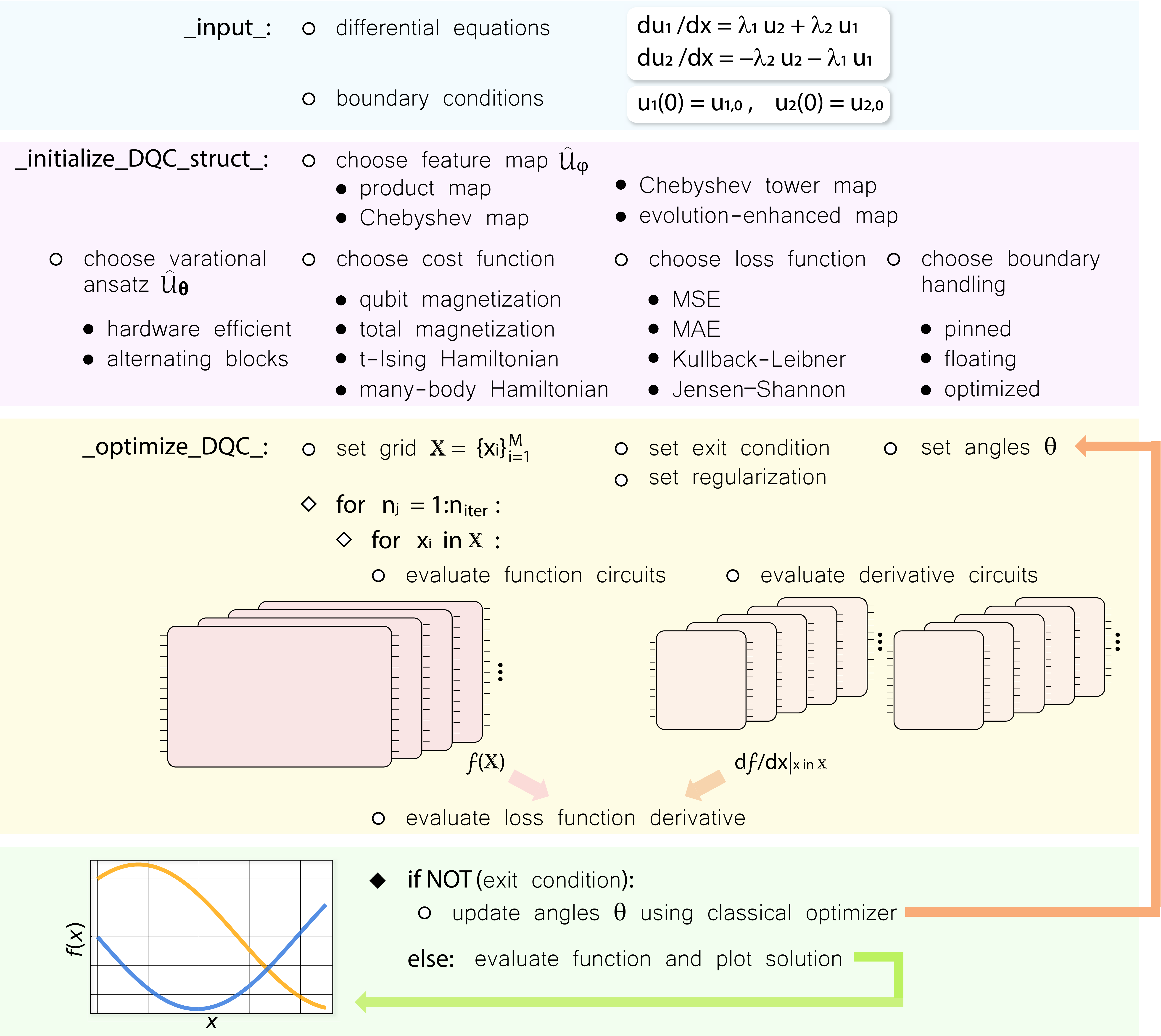}%{fig_workflow_white.eps}
\caption{\textbf{DQC optimization workflow.} The problem is set as a system of differential equations for functions $\bm{u}$, variables $\bm{x}$, and specified boundary conditions. The derivative quantum circuit is constructed by choosing the encoding circuit and optimization schedule. The solution is optimized by evaluating the function and derivative circuits at the defined grid of points $\mathbb{X}$ (which may be multidimensional), and using these values to calculate the loss function derivative. The variational angles are updated in the hybrid quantum-classical loop until a specified exit condition is reached.}
\label{fig:workflow}
\end{figure*}
%%%

In the classical computational methods, the advanced methods for solving differential equations are also currently developed based on machine learning techniques. The solutions are encoded by deep neural networks as universal function approximators~\cite{Sirignano2018}, advancing on the original steps with shallow artificial neural networks~\cite{Lagaris1998}. The approach relies on training physically-informed neural networks~\cite{Raissi2019,Raissi2020}, Fourier neural operators~\cite{ZongyiLi2020} or universal differential equation networks~\cite{Rackauckas2020}. Recently, it has become a part of scientific machine learning ecosystem~\cite{Rackauckas2019}. Another intriguing direction of contemporary research in differential calculus is representation of solutions using tensor networks \cite{Bachmayr2016}. This was also considered as a quantum-inspired and quantum approach to provide speedup for certain problems~\cite{Garcia-Ripoll}.

We propose to solve nonlinear differential equations (DEs) of a general form in a radically different way, where we construct solutions using \emph{differentiable quantum circuits} (or equivalently, \emph{derivative} quantum circuits) that we refer to as `DQCs' in the following \cite{patent}. These can be viewed as quantum neural network circuits that are designed to deal with functions and their derivatives using automatic differentiation rules. The overview of our approach is outlined in Fig.~\ref{fig:workflow}. It relies on defining the DQC structure, choosing the optimization strategy, and training of the circuit parameters that allows for reproducing the solution function. The required elements and workflow are detailed in the next section.

Our approach targets near-term quantum devices, where hybrid quantum-classical computing can work with noisy operation and reasonably shallow circuit depth. We alleviate the data input problem, as we do not rely on amplitude encoding, but use a latent space representation defined in the high-dimensional Hilbert space. This brings advantage in finding solutions of differential equations, as we use the large expressive power of quantum feature maps and parametrized quantum circuits. For efficient readout, we project the high-dimensional function representation to its scalar values by measuring the expectation of an observable (cost operator), and separate evaluation of quantum circuits, offering a NISQ-friendly implementation. We note that our approach also can work with quantum kernels \cite{SchuldPRL}, an advantageous choice for the loss function that allows for efficient multipoint training. However, this approach is more suitable for future highly-coherent quantum devices, and will be considered in forthcoming studies.

%%%===================================

\section{DQC-based differential equations solver: general overview}

\begin{figure*}
\includegraphics[width=1.0\linewidth]{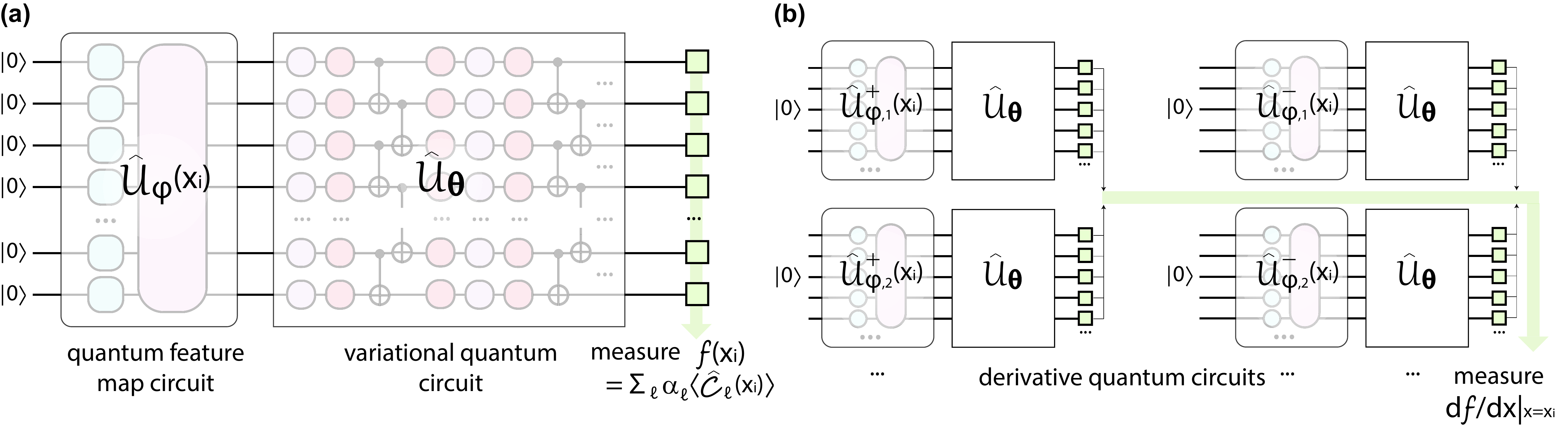}
\caption{\textbf{Differentiable quantum circuits.} (a) Quantum circuit used for encoding the value of a function at a specific value of the variable $x = x_i$. The circuit consists of a feature map $\hat{U}_{\varphi}$ that encodes the $x$-dependence, followed by variational ansatz $\hat{U}_{\bm{\theta}}$, and an observable-based readout for the set of operators $\hat{C}_{\ell}$. The measurement result is classically post-processed to provide a quantum function representation $f(x)$ as a sum of expectations, where coefficients $\alpha_{\ell}$ can be optimized in a quantum-classical hybrid loop. To compose the loss function circuit measurements for different points of optimization grid $\mathbb{X}$ are required. (b) Derivative of the sought function $f(x)$ evaluated at specific point $x = x_i$ is estimated as a sum of expectations for derivative quantum circuits. The full structure follows from the feature map differentiation described in the text and shown for example in Fig.~\ref{fig:feature-map}.}
\label{fig:circuit}
\end{figure*}

We start with an overview description of the differential equation solver based on derivative circuits.
To solve DEs, we prepare trial solutions of the differential equation(s) as quantum circuits parametrized by a variable $x \in \mathbb{R}$ (or a collection of $v$ variables, $\bm{x} \in \mathbb{R}^v$). As the discussion is generalized straightforwardly to the case of $v$ variables, for brevity we henceforth use the simplified single variable notation $x$. We use a \emph{quantum feature map} circuit $\hat{\mathcal{U}}_{\varphi}(x)$ to encode the pre-defined \emph{nonlinear} function of variables $\varphi(x)$ to amplitudes of the quantum state $\hat{\mathcal{U}}_{\varphi}(x) | \text{\O} \rangle$ prepared from some initial state $| \text{\O} \rangle$. A quantum feature map represents a latent space encoding that, unlike amplitude encoding, does not require access to each amplitude and is controlled by classical gate parameters, mapping real parameter $x$ to the corresponding variable value. Sometimes this is also called a quantum \emph{embedding} \cite{Lloyd2020,Schuld2020}, referring to the way data is embedded in the circuit. This follows the steps of quantum circuit learning (QCL) \cite{Mitarai2019}, shown to work well for solving problems of regression and classification. Next, we add a variational quantum circuit $\hat{\mathcal{U}}_{\bm{\theta}}$ parametrized by a vector of variational parameters $\bm{\theta}$ that can be adjusted in a quantum-classical optimization loop. The resulting state $|f_{\varphi,\bm{\theta}}(x)\rangle = \hat{\mathcal{U}}_{\bm{\theta}} \hat{\mathcal{U}}_{\varphi}(x) | \text{\O} \rangle$ for optimal angles contains the $x$-dependent amplitudes sculptured to represent the sought function. Finally, the real valued function can be read out as an expectation value of a predefined Hermitian cost operator $\hat{\mathcal{C}}$, such that the trial function that aims to approximate the solution $u(x)$ reads [Fig.~\ref{fig:circuit}(a)]
\begin{align}
\label{eq:ux}
    f(x) = \langle f_{\varphi,\bm{\theta}}(x) | \hat{\mathcal{C}} |f_{\varphi,\bm{\theta}}(x)\rangle.
\end{align}
%
%%%
%%%

One crucial step of our algorithm is the differentiation of the quantum feature map circuit, $d \hat{\mathcal{U}}_{\varphi}(x) / dx = \sum_j \hat{\mathcal{U}}_{d\varphi, j}(x)$, where the action differential can be represented as a sum of modified circuits $\hat{\mathcal{U}}_{d\varphi, j}$. This allows function derivatives to be represented using the product derivative rule. In the case of quantum feature map generated by strings of Pauli matrices or any involutory matrix we can use the \emph{parameter shift rule} \cite{Mitarai2019,Schuld2019} such that the function derivative is expressed as a sum of expectations [Fig.~\ref{fig:circuit}(b)]
\begin{align}
\label{eq:dudx}
    df(x)/dx = &\frac{1}{2} \sum_j \Big( \langle f_{d\varphi,j,\bm{\theta}}^{+}(x) | \hat{\mathcal{C}} | f_{d\varphi,j,\bm{\theta}}^{+}(x)\rangle \\ \notag &- \langle f_{d\varphi,j,\bm{\theta}}^{-}(x) | \hat{\mathcal{C}} | f_{d\varphi,j,\bm{\theta}}^{-}(x)\rangle \Big),
\end{align}
with $|f_{d\varphi,j,\bm{\theta}}^{\pm}(x)\rangle$ defined through the parameter shifting, and index $j$ running through the individual quantum operations used in the feature map encoding. Applying the parameter shift rule once again we can get the second-order derivative $d^2 f(x)/dx^2$ with four shifted terms for each generator \cite{Mitarai2020,Mari2020}. Importantly, as described we use the automatic differentiation (AD) technique to perform quantum circuit differentiation. AD allows the function derivative to be represented by the exact analytical formula using a set of simple computational rules, as opposed to numerical differentiation. Since automatic differentiation provides an analytical derivative of the circuit at any value of variable $x$, our scheme does not take on the accumulated error from approximating the derivatives. Notably, typical schemes for quantum ODE solvers involve numerical differentiation using Euler's method and finite difference schemes that suffer from approximation errors, and often require a fine discretization grid. This problem is alleviated in our approach.

Our goal is to construct and define the conditions for the quantum circuit to represent the solution of differential equations, generally written as 
\begin{equation}
F[\{d^m f_n/dx^m \}_{m,n}, \{f_n(x) \}_n, x] = 0,
\end{equation} 
where the functional $F[\cdot]$ is provided by the problem for function derivatives of different order $m$ and function/variable polynomials of varying degree $n$. For simplicity we refer to $f$ as a function, and show that the same analysis holds for a vector of functions. The condition above demands that derivatives and nonlinear functions give a net zero contribution. Thus, we can rewrite the task as an optimization problem and use the loss function $\mathcal{L}_{\bm{\theta}}[d_{x}f, f, x]$. This corresponds to minimization of $F[x]|_{x \rightarrow x_i}$ at points in the set $\mathbb{X} = \{ x_i \}_{i=1}^{M}$, and additionally ensuring that the boundary conditions are satisfied. Once the optimal angles 
\begin{align}
    \bm{\theta}_{\mathrm{opt}} = \underset{\bm{\theta}}{\mathrm{argmin}}  (\mathcal{L}_{\bm{\theta}}[d_{x}f, f, x])
\end{align}
are found, we can reproduce the solution from Eq.~\eqref{eq:ux} as a function $f(x)|_{\bm{\theta} \rightarrow \bm{\theta}_{\mathrm{opt}}} \approx u(x)$.

Summarizing the high level description of the proposed approach we note that we: 1) use quantum feature map encoding, thus overcoming the complexity of amplitude encoding for preparing the solution at the boundary; 2) perform automatic differentiation of the quantum feature map circuit, allowing us to represent function derivatives without the imprecision error characteristic to numerical differentiation (finite differencing); 3) search for the suitable solution in the exponential space of fitting polynomials, thus resembling the spectral and finite element methods with improved scaling; 4) avoid the data readout problem, as the solution is encoded in the observable of an operator, such that expectation can be routinely calculated. For the latter point, it differs from amplitude encoding $|u\rangle$ in HHL and related methods, where getting the full solution from amplitudes is exponentially costly and requires tomographic measurements. %Feature space encoding is a way to do it naturally, as we operate observables from the very beginning.
Finally, we note that our goal is to construct circuits that can work for quantum processors with limited computational power, meaning the gate depth (number of operations to performed in series) is limited to a certain amount. This largely defines the training procedure, where we rely on the classical optimization loop. Once deep circuits can be implemented, we can also exploit parallel training strategies for the quantum circuit and quantum state encoding, coming closer to the ideal quantum operation regime.

%================

\section{Methods}

Below we discuss the set of tools that are required to build a differentiable circuit as a solution of differential equations. This corresponds to the main ingredients of the circuit with Sec.~III~\textbf{A} describing quantum feature maps and their derivatives; Sec.~III~\textbf{B} variational quantum circuits (ansatze); Sec.~III~\textbf{C} cost functions; and Sec.~III~\textbf{D} loss functions for the optimization loop. Additionally we detail proposed boundary handling techniques in Sec.~III~\textbf{E}, possible strategies for regularization in Sec.~III~\textbf{F}, encoding multiple functions in Sec.~III~\textbf{G}, and summarize a complete optimization schedule in Sec.~III~\textbf{H}.
%%%
\begin{figure*}
\includegraphics[width=1.0\linewidth]{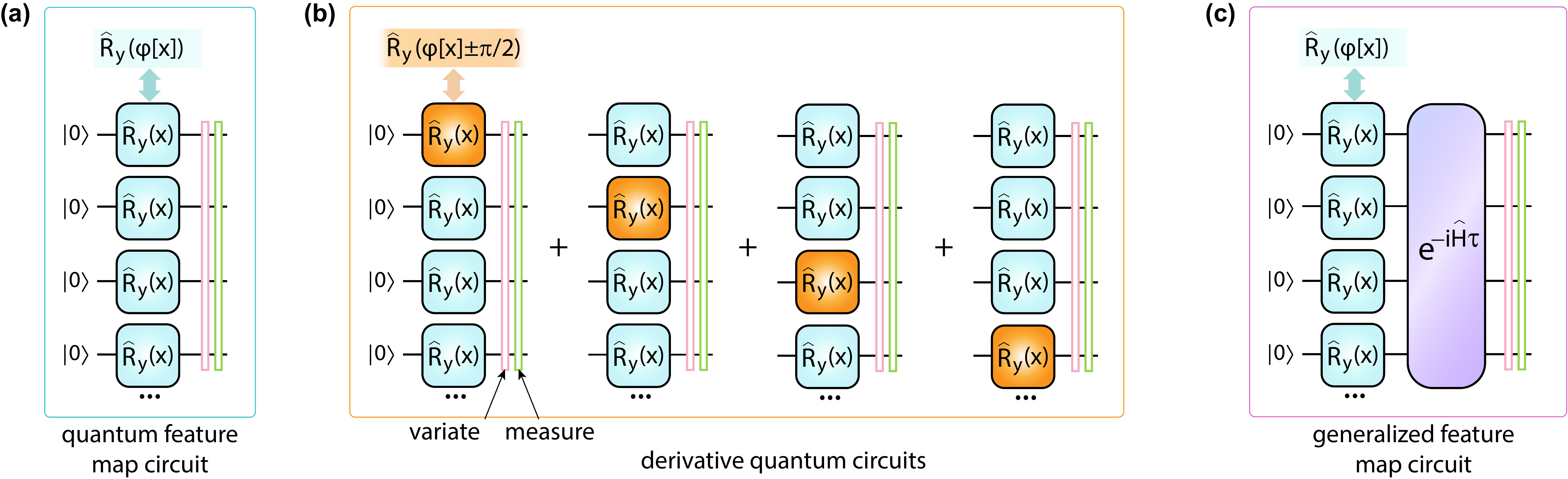}
\caption{\textbf{Product feature map and its derivative.} (a) Quantum feature map of a product type, where single qubit rotations (here chosen as $\hat{R}_y$) act at each qubit individually and are parametrized by a function of variable $x$. Specifically, the expectation value of the circuit is shown, with thin pink and green blocks depicting the variational ansatz and the cost function measurement respectively. For the nonlinear feature encoding the nonlinear function $\varphi(x)$ is used as an angle of rotation. The product feature map can be further generalized to several layers, and different functions $\{ \bm{\varphi} \}$. Several feature maps can be concatenated to represent a multivariable function. (b) shows the derivative quantum circuit for the product feature map. Differentiation over variable $x$ follows the chain rule, with the expectation value of the derivative written as a sum of separate expectations with shifted phases, repeated for each $x$-dependent rotation. (c) Generalized product feature map, which we refer to as \emph{evolution-enhanced feature map}, where the layer of rotation is followed by the unitary evolution generated by Hamiltonian $\hat{H}$. For complicated multiqubit Hamiltonians the encoded state comprises of exponentially many unique $x$-dependent amplitudes. Time interval $\tau$ can be set variationally, or annealed from zero to finite value during the optimization procedure.}
\label{fig:feature-map}
\end{figure*}
%%%

\subsection{Quantum feature maps and their derivatives}

A quantum feature map is a unitary circuit $\hat{\mathcal{U}}_{\varphi}(x)$ that is parametrized by the variable $x$ and typically nonlinear function $\varphi(x)$. Acting on the state it maps $x$ to $\hat{\mathcal{U}}_{\varphi}(x) |\text{\O}\rangle$ such that the $x$-dependence is translated into quantum state amplitudes~\cite{BenedettiRev}. This is also referred to as a latent space mapping. There exist different ways of feature map encoding. Below we describe some possible choices and propose a distinct \emph{Chebyshev quantum feature map} that allows the approximation of highly nonlinear functions. We also describe the procedure of feature map differentiation, the crucial step of constructing quantum circuits for solutions of differential equations.\\

\textbf{Product feature maps.} First, we consider a product feature map that uses qubit rotations, but has nonlinear dependence on the encoded variable $x$. In the simplest case this corresponds to a single layer of rotations written in the form
\begin{align}
\label{eq:feature_QCL}
\hat{\mathcal{U}}_{\varphi}(x) = \bigotimes_{j=1}^{N'} \hat{R}_{\alpha,j} (\varphi[x]),
\end{align}
where $N' \leq N$ is the number of qubits used for the encoding. $\hat{R}_{\alpha,j}(\varphi) := \exp\left(-i \frac{\varphi}{2} \hat{P}_{\alpha,j} \right)$ is a Pauli rotation operator for Pauli matrices $\hat{P}_{\alpha,j} = \hat{X}_j$, $\hat{Y}_j$, or $\hat{Z}_j$ ($\alpha = x,y,z$, respectively) acting on qubit $j$ with phase $\varphi$. This represents the feature map choice used in quantum circuit learning \cite{Mitarai2019}, and several similar encodings were discussed in \cite{Vidal2019, Romero2019}, and reviewed in \cite{BenedettiRev}. The next step is to assign a nonlinear function for rotation, with a popular choice being $\varphi(x) = \arcsin x$ and $\alpha = y$ such that only real amplitudes are generated. The corresponding circuit is shown in Fig.~\ref{fig:feature-map}(a). The unitary operator Eq.~\eqref{eq:feature_QCL} is then rewritten as
\begin{align}
\label{eq:feature_L1}
\hat{\mathcal{U}}_{\varphi}(x) = \bigotimes_{j=1}^N \exp\left(- i \frac{\arcsin x}{2} \hat{Y}_j \right),
\end{align}
leading to amplitudes that depend on the encoded variables as $\cos[ (\arcsin x)/2]$ and $\sin[ (\arcsin x)/2]$. Acting on the initial state $|\text{\O} \rangle$ this feature map encodes the variable as an $N$-th degree polynomial formed by $\{1, x, \sqrt{1-x^2} \}$ and their products \cite{Mitarai2019}. The redundancy from many qubits thus forms a basis set for function fitting \cite{Vidal2019}.

The product feature map can be generalized to several layers of rotations $\ell = 1, 2, ..., L$, various nonlinear functions $\varphi_\ell$ and specific subsets of qubits $\mathbb{N}$, written as
\begin{align}
\label{eq:feature_L1_multi}
\hat{\mathcal{U}}_{\varphi}(x) = \prod_{\ell=1}^L \bigotimes_{j \in \mathbb{N}_\ell} \hat{R}_{\alpha,j}^{(\ell)} (\varphi_\ell[x]).
\end{align}

Next, we show how the quantum feature map can be differentiated. Let us use the example in Eq.~\eqref{eq:feature_QCL} considering $\alpha = y$ rotations and a full layer. The derivative for a unitary operator generated by an involutory matrix (length-1 Pauli string in this case) can be written as
\begin{align}
\label{eq:feature_L1_derivative}
\frac{d}{dx} \hat{\mathcal{U}}_{\varphi}(x) = \frac{1}{2} \left( \frac{d}{dx} \varphi(x) \right) \sum_{j' = 1}^N \bigotimes_{j=1}^{N} (-i \hat{Y}_{j'} \delta_{j,j'}) \hat{R}_{y,j} (\varphi[x]) 
\\ \notag = \frac{1}{2} \left( \frac{d}{dx} \varphi(x) \right) \sum_{j' = 1}^N \bigotimes_{j=1}^{N} \hat{R}_{y,j} (\varphi[x] + \pi \delta_{j,j'}) ,
\end{align}
where Euler's formula can be used to rewrite the derivative into the form of a sum of unitaries, where $x$-dependent rotations are shifted one-by-one. Next, the formula can be generalized to the expectation value of any observable $\langle \hat{\mathcal{C}} \rangle$ for the encoded state, following the steps of a standard parameter shift rule \cite{Mitarai2019,Schuld2019}. This reads
\begin{align}
\label{eq:C_derivative}
\frac{d}{dx} \langle \text{\O} | \hat{\mathcal{U}}_{\varphi}(x)^\dagger  \hat{\mathcal{C}} \hat{\mathcal{U}}_{\varphi}(x) |\text{\O} \rangle = \frac{1}{4} \left( \frac{d}{dx} \varphi(x) \right) \left( \langle \hat{\mathcal{C}} \rangle^{+} - \langle \hat{\mathcal{C}} \rangle^{-} \right)
\end{align}
where $\langle \hat{\mathcal{C}} \rangle^{+}$ and $\langle \hat{\mathcal{C}} \rangle^{-}$ are the sums of shifted unitaries
\begin{align}
\label{eq:C_pm}
\langle \hat{\mathcal{C}} \rangle^{\pm} = \sum_{j' = 1}^N \bigotimes_{j=1}^{N} \langle \text{\O} | \hat{R}_{y,j}^{\dagger} (\varphi[x] \pm \frac{\pi}{2} \delta_{j,j'}) \hat{\mathcal{C}} \hat{R}_{y,j} (\varphi[x] \pm \frac{\pi}{2} \delta_{j,j'}) |\text{\O} \rangle.
\end{align}
The corresponding derivative quantum circuits \eqref{eq:C_derivative} are shown in Fig.~\ref{fig:feature-map}(b), where differentiation of the cost function for feature map is performed using the chain rule (highlighted rotations).  A similar strategy can be applied for generic multilayer feature maps and a different choice of nonlinear map $\varphi(x)$. Finally, in the cases where the generator of the feature map (encoding Hamiltonian $\hat{H}$) is not an involutory matrix we can rewrite them as a sum of unitary operators, and measure the derivative as a sum of overlap measurements using the SWAP test \cite{Mitarai2020,MitaraiPRRes,Kyriienko2020}.\\

%---

\textbf{Chebyshev feature maps.} Next, we consider a distinct choice of nonlinear quantum feature map that we name the \emph{Chebyshev feature map}. Belonging to the product feature map family, it drastically changes the basis set for function representation. As a building block we use a single qubit rotation $\hat{R}_{y,j}(\varphi[x])$, but with nonlinearity introduced as $\varphi(x) = 2 n \arccos x$, $n = 0,1,2,..$, such that the encoding circuit reads
\begin{align}
\label{eq:Cheb-feat-general}
\hat{\mathcal{U}}_{\varphi}(x) = \bigotimes_{j=1}^{N} \hat{R}_{y,j}(2  n[j] \arccos x).
\end{align}
Here we consider that the coefficient $n[j]$ may in general depend on the qubit position $j$. Note that the seemingly small change of factor two multiplication for $\varphi(x) = 2 \arccos x$ (as compared to previously considered product map with $\varphi(x) = \arccos x$) goes a surprisingly long way. Namely, let us expand the rotation using Euler's formula, getting
\begin{align}
\label{eq:Ry-to-Cheb}
    &\hat{R}_{y,j}(\varphi[x]) = \exp\left(-i \frac{2 n \arccos(x)}{2} \hat{Y}_j\right) \\
    \notag &= \cos(n \arccos(x)) \mathbbm{1}_j - i \sin(n\arccos(x)) \hat{Y}_j \\ \notag &= T_n(x) \mathbbm{1}_j + \sqrt{1-x^2} U_{n-1}(x) \hat{X}_j \hat{Z}_j .
\end{align}
The resulting decomposition \eqref{eq:Ry-to-Cheb} corresponds to a unitary operation with matrix elements defined by degree-$n$ Chebyshev polynomials of first and second kind, denoted as $T_n(x)$ and $U_n(x)$, respectively \cite{AbramowitzStegun}. 
The low degree Chebyshev polynomials of the first kind are written explicitly as $T_0(x) = 1$, $T_1(x) = x$, $T_2(x) = 2x^2 - 1$, $T_3(x) = 4x^3 - 3x$, and higher degrees can be deduced using the recursion relation
\begin{align}
\label{eq:Chebyshev_Tn_recursion}
    T_{n+1}(x) = 2 x T_n(x) - T_{n-1}(x).
\end{align}
Similarly, we can write second-kind Chebyshev polynomials as $U_0(x) = 1$, $U_1(x) = 2x$, $U_{n+1}(x) = 2xU_n(x) - U_{n-1}(x)$. The crucial properties of Chebyshev polynomials are their chaining properties, nesting properties, and simple differentiation rules. The chaining properties for polynomials of the first and second kind read as $2 T_m (x) T_n (x) = T_{m+n}(x) + T_{|m-n|}(x)$ and $U_m (x) U_n (x) = \sum_{k=0}^n U_{m-n+2k}(x)$, respectively. Derivatives can be obtained as $d T_n(x)/dx = n U_{n-1}(x)$. Nesting corresponds to the relation $T_n (T_m (x)) \equiv T_{nm}(x)$. Finally, polynomials of different kinds can be converted between as $U_n(x) = 2 \sum_{j \text{ even}}^n T_j(x)$ when $n~\mathrm{is~even}$, and $U_n(x) = 2 \sum_{j \text{ odd}}^n [T_j(x) - 1]$ when $n~\mathrm{is~odd}$. Finally, we note that Chebyshev polynomials represent oscillating functions defined in the region $x=(-1,1)$, and their derivatives diverge at the boundaries of this interval.
%%%
\begin{figure}
\includegraphics[width=1.0\linewidth]{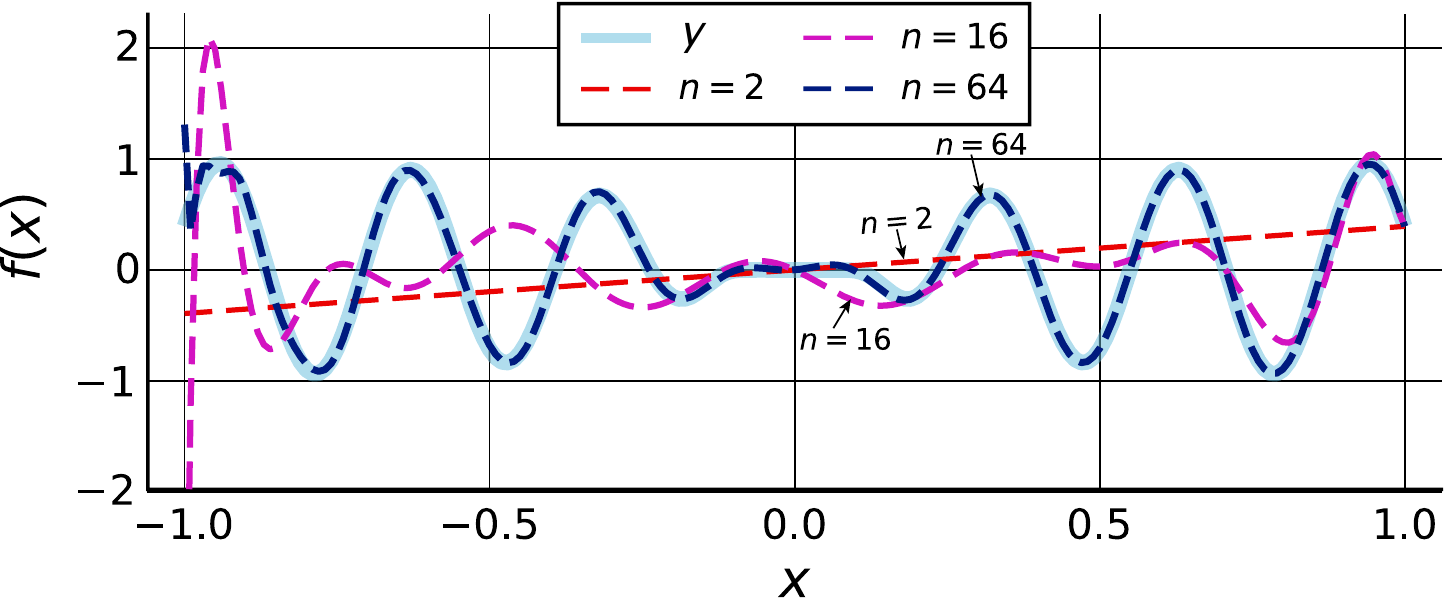}
\caption{\textbf{Fitting with Chebyshev polynomials.} As an example we show a fit of the function $y(x) = \exp(-1/25x^2)\text{cos}(20x)$ using different number $n$ Chebyshev polynomials up to degree $n-1$. We note here that the target function is oscillatoric and is decaying exponentially. The thick solid curve denotes the desired function $y(x)$ and the dashed curves are the obtained fitting curves (see legend). Their fit improves with $n$. The basis set with $n=2$ is unable to represent the function faithfully (only the linear fit is available). Increased basis set size with $n=16$ represents the desired function in some regions of $x$, and $n=64$ successfully provides an accurate fit.}
\label{fig:cheb-fourier-fitting}
\end{figure}
%%%

The power of the described representation can be inferred from approximation theory. It states that any smooth function can be represented as $f(x) = \sum_{n=0}^{\infty} A_n T_n(x), \quad |x|\leq 1$. An example of fitting a set of Chebyshev polynomials of the first kind to a function is shown in Fig. \ref{fig:cheb-fourier-fitting}. As expected the more basis functions we use, the more accurate of a fit we can obtain. As shown in the Remez algorithm \cite{RemezUkrPhys,Powell1981}, Chebyshev polynomials form an optimal set of basis functions in the sense of the uniform $L_\infty$ norm. This is why they are at the foundation of spectral algorithms for solving ODEs \cite{Childs2020a}, and also give an edge in quantum simulation \cite{Berry2015,Childs2017}.

In the present study we consider two types of Chebyshev quantum feature maps. The first version corresponds to a \emph{sparse Chebyshev feature map} defined as
\begin{align}
\label{eq:Cheb-feat-sparse}
\hat{\mathcal{U}}_{\varphi}(x) = \bigotimes_{j=1}^{N} \hat{R}_{y,j}(2 \arccos x),
\end{align}
where the encoded degree is homogeneous and equal to one. Here we make use of the chaining properties $T_n(x)$ and $U_n(x)$, noting that once we create states with Chebyshev polynomials as prefactors, the basis set will grow further by concatenating elements. Henceforth we drop the \emph{sparse} distinction and simply refer to \eqref{eq:Cheb-feat-sparse} as \emph{Chebyshev feature map}. Importantly, the unitary operators in Eq.~\eqref{eq:Cheb-feat-sparse} have nonlinear dependence on variable $x$, leading to a harmonic feature mapping.

The second version we consider corresponds to a \emph{Chebyshev tower feature map} defined as
\begin{align}
\label{eq:Cheb-feat-tower}
\hat{\mathcal{U}}_{\varphi}(x) = \bigotimes_{j=1}^{N} \hat{R}_{y,j}(2 j \arccos x),
\end{align}
where the encoded degree grows with the number of qubits, creating a tower-like structure of polynomials with increasing $n=j$. Again, as polynomials chain together and morph between two kinds and their degrees, the basis set is largely enriched. This is the choice we exploit when large expressibility is needed without increasing system size and number of rotations. Eq.~\eqref{eq:Cheb-feat-tower} allows the representation of generic functions, and can be improved further by using layers of rotations as in Eq.~\eqref{eq:feature_L1_multi}.

\textbf{Evolution-enhanced feature maps.} Product feature maps induce nonlinear mappings between variable(s) $x$ and quantum states described by tensor products of separate single-qubit wavefunctions. These states are limited to the subspace of product states. To utilize the power of the entire Hilbert space of the system, approaching the amplitude encoding case, we need to populate independently distinct amplitudes, including the subspace of entangled states. To make the described feature maps even more expressive, we suggest enhancing product feature maps (and specifically the layered Chebyshev map) with additional entangling layers represented by Hamiltonian evolution. Namely, after the set of single qubit rotations we consider another unitary $\exp(-i \hat{H} \tau)$ which acts for time $\tau$ and is generated by the Hamiltonian $\hat{H}$. The sketch of the circuit is shown in Fig.~\ref{fig:feature-map}(c). By choosing $\hat{H}$ as a complex many-body Hamiltonian we ensure that exponentially many amplitudes are generated. It is known that the quantum simulation of dynamics leads to a volume-like increase of entanglement. %[TensorNets]. 
One important choice is when $\hat{H}$ corresponds to a hard problem from NP-hard complexity class, as proposed in Ref.~\cite{Havlicek2019}. Then using two layers of rotations plus evolution the embedding becomes difficult to simulate classically, but can be implemented as a unitary evolution on a quantum computer. Additionally, we envisage an evolution layer that is parameter-dependent, $\tau(x)$. In this case the \emph{evolution-enhanced feature map} can also be seen through the prism of the recently proposed Fourier feature maps \cite{Schuld2020}. This class of quantum feature maps is based on the evolution operator $\exp(-i \hat{H}_{\mathrm{data}} x)$, which is applied for some set of qubits. The Fourier map lets functions be encoded as Fourier series defined by the differences of the eigenvalues of $\hat{H}_{\mathrm{data}}$. As it involves unitary operators with phases being linear functions of $x$, this is a fully harmonic mapping. The evolution-enhanced feature map then joins the Chebyshev and Fourier basis sets, encoded in the full Hilbert space for complex $\hat{H}$.

\textbf{Digital quantum feature maps.} Another possibility for encoding the data using a feature map is to transform a data instance into a computational basis state, which we refer to as a \textit{digital quantum feature map}. This relates $x$, written in binary form, to the corresponding state $|x\rangle$ in binary representation. The feature map circuit $\hat{\mathcal{U}}_{\varphi}(x)$ used to encode the binary variable $x$ reads $\hat{\mathcal{U}}_{\varphi}(x) = \bigotimes_{j=1}^{N} \exp(-i \frac{\pi}{2} x_j \hat{X}_j)$, where $\{ x_j \}$ denote binary values for the parameter $x$ in $j$-th digit. The differentiation of the digital feature map then relies on the product rule for $N$ rotations, and also includes the binary derivative of the variable from the product rule. 

Another possibility is converting the variable into a decimal representation as $x_{\mathrm{int}} = \sum_j x_j \cdot 2^j $. For the reverse procedure we can identify each binary digit $x_j$ by the remainder of the repeated division $x_j = \mathrm{mod}(x_{\mathrm{int}}, 2^j)$. We can thus rewrite $\hat{U}_{\varphi}(x)$ as a function of $x_{\mathrm{int}}$, and learn how to differentiate circuits with this feature map with respect to $x = x_{\mathrm{int}}$. 

Digital quantum feature maps offer a potentially powerful technique when dealing with functions of discrete variables. Paired with expressive variational ansatze they may offer function encoding with memory savings of the amplitude encoding, while avoiding the input problem. The details of digital feature map training will be discussed in future works.

\subsection{Variational quantum circuits}

To construct the solution of differential equations as a quantum circuit we need to manipulate the latent space basis function and bring both derivatives and function to the required form. This is achieved through the variational circuit $\hat{\mathcal{U}}_{\bm{\theta}}$, typically referred to as a \emph{variational quantum ansatz}. Below we detail architectures employed.
%%%
\begin{figure}
\includegraphics[width=0.9\linewidth]{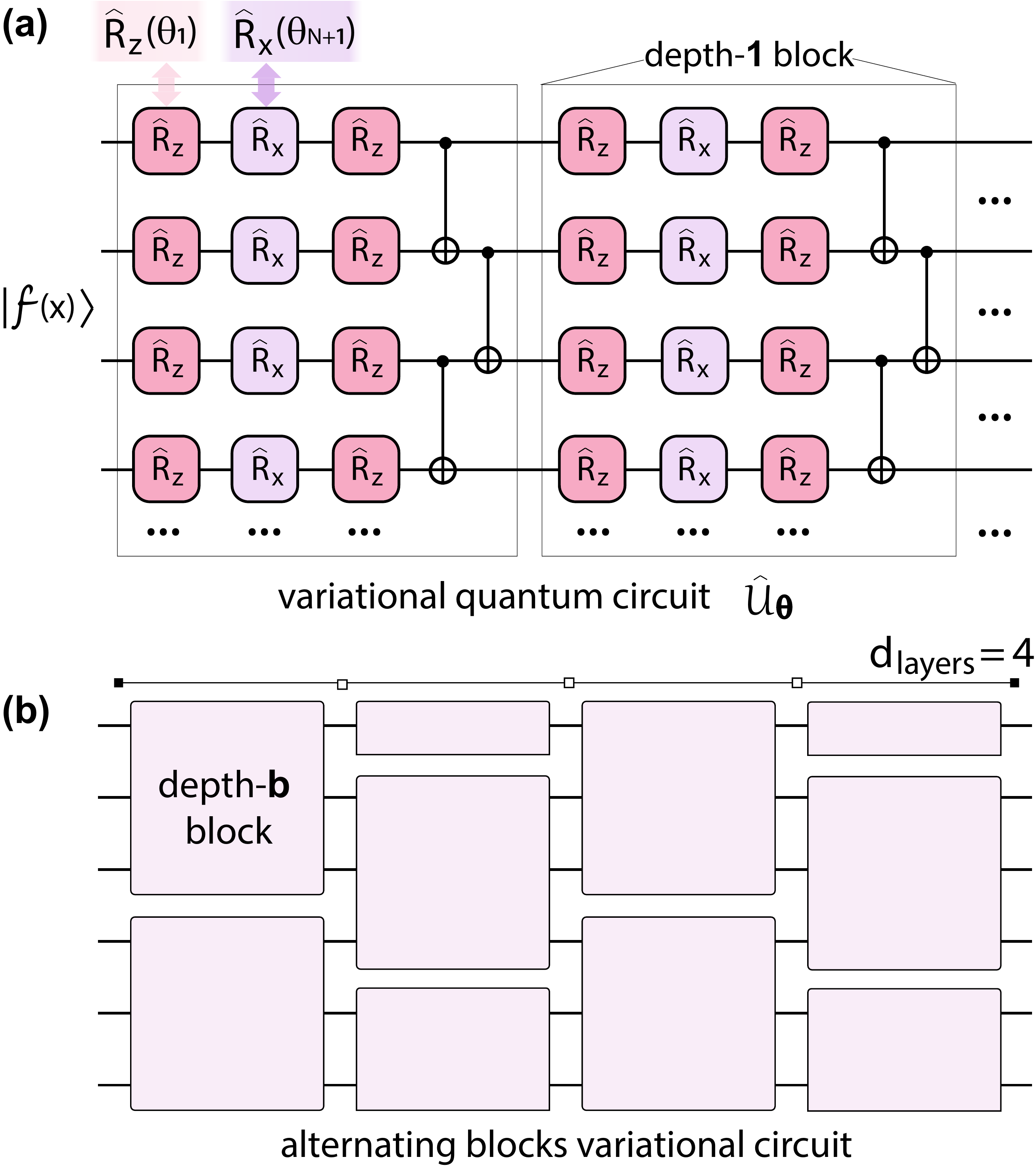}
\caption{\textbf{Variational quantum circuits.} (a) A variational ansatz in the hardware-efficient form. It consists of a parametrized rotation layer forming $\hat{R}_z$-$\hat{R}_x$-$\hat{R}_z$ pattern, such that an arbitrary single qubit rotation can be implemented. Variational angles $\bm{\theta}$ are set for each rotation individually. The rotation layer is then followed by an entangling layer chosen as CNOT operations between nearest neighbours. The blocks of ``rotations-plus-entangler'' are repeated $d$ times to form the full variational circuit $\hat{\mathcal{U}}_{\bm{\theta}}$. (b) Alternating blocks ansatz. The variational circuit consists of blocks of width $N_b$ qubits ($N_b/2$ for boundary qubits). Blocks are chosen in the hardware-efficient form shown in (a) with depth of $b$. The blocks are placed in a checkerboard pattern, and repeated $d_{\mathrm{layers}}$ times. The goal is to entangle qubits locally, while avoiding global entangling operations that can result in vanishing gradients during $\bm{\theta}$ optimization.}
\label{fig:ansatz}
\end{figure}
%%%

\textbf{Hardware efficient ansatz.} As a first choice of variational circuit $\hat{\mathcal{U}}_{\bm{\theta}}$ we consider layers of parametrized rotations, followed by layers of CNOT operations. This is known as a hardware efficient ansatz (HEA) which was originally proposed for VQE for chemistry applications \cite{Kandala2017}. The structure of a HEA quantum circuit corresponds to concatenated layers of single qubit rotations and global entangling layers for all $N$ qubits, shown schematically in Fig.~\ref{fig:ansatz}(a). Rotations are arranged in a $\hat{R}_z$-$\hat{R}_{x}$-$\hat{R}_{z}$ sequence parametrized by independent angles $\bm{\theta}$ such that arbitrary single-qubit operations can be reproduced. The entangling layer is chosen as a network of CNOTs. We here consider specifically a linear quantum device connectivity, but the entangling layer can be generalized to other connectivities. The block of rotations plus CNOTs is then repeated for a depth of $d$ times. As the number of layers $d$ grows, the circuit's expressive power increases. Expressive power is the ability to represent arbitrary $N$-qubit unitary gates. However, this comes with an increased number of controlled parameters which can complicate the search for an optimal $\bm{\theta}_{\mathrm{opt}}$ for the solution, known as a problem in trainability. This can change depending on the variational ansatz type and cost function choice \cite{Cerezo2020}. 

\textbf{Alternating blocks ansatz.} A second option is to use the alternating blocks ansatz (ABA), where instead of global entangling layers separate subblocks are used, interleaved into a checkerboard form [Fig.~\ref{fig:ansatz}(b)]. Each subblock has a hardware efficient form shown in Fig.~\ref{fig:ansatz}(a) for the specified depth $b$. For the first layer the width of the subblock (number of active qubits) is equal to $N_b$ such that $\ceil{N/N_b}$ blocks are used (and is smaller than $N_b$ if $N/N_b$ is not an integer). The next layer consists of the same subblocks, but is shifted by $\floor{N_b/2}$, where subblocks at the ends are adjusted to span the remaining qubits. The described checkerboard-like structure is repeated for $d_{\mathrm{layers}}$. The motivation behind ABA is that we want to entangle qubits locally first, and gradually form a correlated state by interleaving subblocks. This helps to improve trainability of the circuit together while maintaining high expressibility \cite{Cerezo2020, Yamamoto2020}.\\

Here, several points are due. First, we note that the choice of ansatz is sensitive to the choice of the cost function operator (see the next subsection), as dictated by the symmetry. Namely, as a consequence of cost function choice we need to choose non-commuting generators $\hat{\mathcal{G}}_j$ for the variational ansatz such that  $[\hat{\mathcal{C}}, \hat{\mathcal{G}}_j] \neq 0$. This ensures that the solution space can be spanned. Also, we can account for the symmetries, reducing the Hilbert space for the solution search. In many cases generators can be chosen such that only real amplitudes are generated, see for instance~\cite{Cerezo2020}. We can also use adaptive strategies or a genetic search~\cite{Grimsley2019,Tang2019,Chivilikhin2020}.

Second, to search for the optimal circuit parameters we use stochastic gradient descent, and specifically its adaptive version represented by {\sffamily{}Adam} \cite{adam}. For this, the gradients of the variational circuit $\nabla_{\bm{\theta}} \hat{\mathcal{U}}_{\bm{\theta}}$ are measured using the automatic differentiation approach, as performed in previous studies \cite{Mitarai2019,Schuld2019,Zeng2019}. The gradient is also weighted with a learning rate, commonly referred to as $\alpha$. Choosing an ansatz parametrized by single-qubit rotations allows the application of the parameter shift rule, while overlap measurement opens up options for more general strategies \cite{MitaraiPRRes}.

Finally, we note that choosing the optimal ansatz is an open challenge for improving DQC training. In comparison to chemistry problems where the structure may be inspired by physical reasoning \cite{OxfordRev,ZapataRev}, this is not the case for differential equation solvers and further studies in this direction are needed.

\subsection{Cost function}

To read out information we choose a Hermitian cost operator $\hat{\mathcal{C}}$ to measure as an observable. This allows relating the expectation of $\hat{\mathcal{C}}$, parametrized by nonlinear variable dependence and variational angles, to the scalar function $f(x)$ as $\langle f_{\varphi,\bm{\theta}}(x) | \hat{\mathcal{C}} |f_{\varphi,\bm{\theta}}(x)\rangle$. In general many possible choices of cost operators are available. The simplest example corresponds to the magnetization of a single qubit $j$, $\langle \hat{Z}_j \rangle$. Note that this choice allows representing functions in range $[-1,1]$, and requires rescaling for other intervals. Other choices include total magnetization in the system $\hat{C} = \sum_j \hat{Z}_j$ with equal or randomized weights.

Additionally, we can choose the cost as a quantum Hamiltonian that has a provably complex spectrum, and for instance belongs to ergodic phase. This can be written as an Ising Hamiltonian with additional transverse and longitudinal magnetic fields,
\begin{align}
    \hat{C} = \sum_j J_{j,j+1} \hat{Z}_j \hat{Z}_{j+1} + h_{j}^z \hat{Z}_j + h_{j}^x \hat{X}_j ,
\end{align}
where the Ising couplings $J_{j,j+1}$ and $h_{j}^{z,x}$ can be inhomogeneous. We note that for cost functions with several non-commuting groups of observables we can use a Hamiltonian averaging procedure, where term-by-term measurement is performed. Moving on from nearest-neighbour Hamiltonians, we can also exploit spin-glass type cost functions of the form
\begin{align}
    \hat{C} = \sum_{i<j} J_{i,j} \hat{Z}_i \hat{Z}_j + \sum_j h_{j}^z \hat{Z}_j .
\end{align}
These are known to include NP-hard problem instances, and they allow for high expressibility of the circuit describing the DE's solution. Finally, a generic cost function may contain a large set of Pauli strings, similar to some instances in quantum chemistry. 

Together with measuring individual cost functions, we also consider functions being represented by a classically weighted sum of observables. This reads
\begin{align}
    \hat{C} = \sum_{\ell} \alpha_\ell \hat{C}_\ell,
\end{align}
where $\alpha_\ell \in \mathbb{R}$ are weighting coefficients, and $\hat{C}_\ell$ are cost functions that can be chosen from the pool of operators described above. Importantly, we consider the coefficients $\alpha_\ell$ to be tunable, such that the gradient descent (represented by {\sffamily{}Adam} in our case) can adjust the cost to have optimal form. This procedure further harnesses the strength of the proposed hybrid quantum-classical workflow.

\subsection{Loss function}

To solve the system of differential equations we need to provide a way to quantify how well the DQC-suggested trial function matches the conditions to represent the solution of the problem being considered. The classical optimizer can then update the variational parameters to reduce this `distance'. This distance corresponds to the difference between a differential equation (all terms collected on one side) and zero. The difference is evaluated at a set of points. We also need to check if the solution matches initial and boundary conditions. This can be recast as an optimization problem for a loss function of derivatives and functions evaluated at a grid of points.

We can write a loss function parametrized by variational parameters $\bm{\theta}$ in the general form
\begin{align}
\label{eq:Loss_total}
    \mathcal{L}_{\bm{\theta}}[d_{x}f, f, x] &= \mathcal{L}_{\bm{\theta}}^{\mathrm{(diff)}}[d_{x}f, f, x] + \mathcal{L}_{\bm{\theta}}^{\mathrm{(boundary)}}[f, x],
\end{align}
where we split the loss contribution from matching the differentials $\mathcal{L}_{\bm{\theta}}^{\mathrm{(diff)}}$ and the loss contribution from satisfying the boundary conditions $\mathcal{L}_{\bm{\theta}}^{\mathrm{(boundary)}}$. The differential loss is defined as
\begin{align}
\label{eq:Loss_diff}
    \mathcal{L}_{\bm{\theta}}^{\mathrm{(diff)}}[d_{x}f, f, x] &= \frac{1}{M} \sum_{i=1}^M L\big( F[d_{x}f(x_i), f(x_i), x_i], 0 \big)
\end{align}
with $L(a, b)$ being a function describing how the distance between the two arguments $a$ and $b$ is being measured. The loss is estimated on a grid of $M$ points, and is normalized by the grid size. Functional $F$ corresponds to the differential equation written in the form $F[d_{x}u, u, x] = 0$. It can be evaluated by combining values of $f$ and $d_{x}f$ at the training grid points. We stress that the functional includes information about all differential equations when we deal with the system, such that contributions from all equations are accounted for. The boundary loss contribution reads
\begin{align}
\label{eq:Loss_boundary}
    \mathcal{L}_{\bm{\theta}}^{\mathrm{(boundary)}}[f, x] &= \eta L(f(x_0), u_0)
\end{align}
which includes the distance between the function value at the boundary $x_0$ and given boundary value $u_0$. We note that $x_0$ can be an initial point or a set of boundary points. We also introduce $\eta$ as a \emph{boundary pinning} coefficient that controls the weight of the boundary term in the optimization procedure. In particular, larger $\eta > 1$ may be used to ensure the boundary is prioritised and represented to higher precision.

We have considered several choices of the loss defined by three distance definitions $L$. The first loss type corresponds to the \emph{mean square error} (MSE) introduced as
\begin{align}
\label{eq:MSE}
    L(a,b) = (a-b)^2.
\end{align}
While being simple, we find the choice \eqref{eq:MSE} intuitive and performing sufficiently well in numerical simulations. Additionally we also consider \emph{mean absolute error} (MAE) loss defined with distance $L(a,b) = |a-b|$.
%
%\begin{align}
%\label{eq:MAE}
%    L(a,b) = |a-b|
%\end{align}
%
Finally, several more complex metrics can be used, including variants of Kullback-Leibler (KL) divergence and Jensen–Shannon divergence. Being the loss functions routinely used in statistical modelling, we expect them to perform well for specific systems.
%The modified Kullback-Leibler loss is
%
%\begin{align}
%    L(a,b) = a^2 \text{log}(\frac{a^2 + 1}{b^2+1})
%\end{align}
%

The choice of loss functions dictates how the optimizer perceives the distance between vectors and therefore affects the convergence. MSE places a greater emphasis on larger distances and smaller weight on small distances, strongly discouraging terms with large $L$. Both MAE and KL do not place such an emphasis and may  have slower convergence. However, once close to the optimal solution they can achieve higher accuracy than MSE. %KL has an additional incentive for keeping the magnitude of the first argument low, which for the differential loss term works well as we know we are going to want to match our differential equation to zero. 

\subsection{Boundary handling}

As our goal is to construct a quantum circuit that satisfies a system of differential equations, together with matched derivatives we need to ensure that an initial value or boundary value problem is solved. Generally this corresponds to fixing the function value at a required initial point or a collection of boundary points, thus resembling the quantum circuit learning tasks considered in Ref.~\cite{Mitarai2019}. At the same time, there are several ways how the DQC-based function $f_{\bm{\theta}}(x)$ can be constructed, leading to varying performance and specific pros/cons when solving particular problems.

Information about the boundary can be included as part of the loss function [Eq.~\eqref{eq:Loss_total}]. For the MSE loss function type the boundary part \eqref{eq:Loss_boundary} can be written in the form
\begin{align}
\label{eq:boundary_loss}
    \mathcal{L}_{\bm{\theta}}^{\mathrm{(boundary)}}[f, x] = \eta \Big(f_{\bm{\theta}}(x_0) - u_0 \Big)^2,
\end{align}
where $x_0$ represents the set of boundary points (or an initial point), and $u_0$ is a vector of boundary values, and $\eta$ is a pinning coefficient as described previously.\\

\textbf{Pinned boundary handling.} The first option is to include the information about the boundary in the expectation of the cost function. This corresponds to simply choosing a cost operator $\hat{\mathcal{C}}$, and representing the solution in the form 
\begin{align}
\label{eq:pinned_boundary}
f(x) = \langle f_{\varphi,\bm{\theta}}(x) | \hat{\mathcal{C}} |f_{\varphi,\bm{\theta}}(x)\rangle.
\end{align}
The initial value $u_0$ is then matched via the boundary term in the loss function. The strength of the \emph{pinned boundary} handling is in equivalent treatment of boundary and derivative terms, both being encoded in $\hat{\mathcal{C}}$. At the same time, the weakness corresponds to the necessity of adjusting the boundary value starting from the one represented by initial $\bm{\theta}_{\mathrm{init}}$, typically generated randomly. This can be adjusted by shifting $f(x)$ by a constant-times-identity term added to the cost operator, $\hat{\mathcal{C}} = \alpha_0 \mathbbm{1} + \sum_{j=1}^{M} \alpha_j \hat{\mathcal{C}}_j$, where $\alpha_0$ is set such that for $\bm{\theta}_{\mathrm{init}} \sim \mathrm{random}[0,2\pi]$ the function $\langle f_{\varphi,\bm{\theta}_{\mathrm{init}}}(x) | \hat{\mathcal{C}} |f_{\varphi,\bm{\theta}_{\mathrm{init}}}(x)\rangle$ typically lies close to $u_0$ value when evaluated at $x = x_0$.

\textbf{Floating boundary handling.} The second choice of the boundary handler corresponds to iteratively shifting the estimated solution based on the boundary or initial point. For this method the boundary information does not require a separate boundary loss term nor is it encoded in the expectation of the cost function. Instead it is set iteratively within the parametrization of the function. As the function is parametrized to match a specific boundary, information about the boundary is still contained within the function and its derivatives. Therefore boundary information is still present within the loss function despite there not being a separate boundary loss term. We parametrize the function as
\begin{align}
\label{eq:ux_float}
    f(x) = f_{\mathrm{b}} + \langle f_{\varphi,\bm{\theta}}(x) | \hat{\mathcal{C}} |f_{\varphi,\bm{\theta}}(x)\rangle,
\end{align}
with $f_{\mathrm{b}} \in \mathbb{R}$ being a parameter adjusted after each iteration step as
\begin{align}
\label{eq:ux_float_ub}
    f_{\mathrm{b}} = u_0 -  \langle f_{\varphi,\bm{\theta}}(x_0) | \hat{\mathcal{C}} |f_{\varphi,\bm{\theta}}(x_0)\rangle .
\end{align}
This effectively allows the solver to find a function $\langle f_{\varphi,\bm{\theta}}(x) | \hat{\mathcal{C}} |f_{\varphi,\bm{\theta}}(x)\rangle$ which solves the differential equation shifted to any position, then being shifted to the desired initial condition as shown in Eq.~\eqref{eq:ux_float}. This method of boundary handling guarantees exact matching to initial values given and does not require a separate boundary term in the loss function, thus the derivative loss term does not have to compete with the boundary loss. Furthermore, as we allow the cost function to match to the solution shifted by any amount, this simplifies the choice for optimal angles and removes the dependence on initial $\bm{\theta}_{\mathrm{init}}$. However this method does require knowledge of an exact initial value which can be an issue in specific situations. This technique can be generalized for multivariable problems which have an initial condition as a function of a subset of the independent variables. Evaluation of the derivatives of the initial condition function is required for encoding the partial derivatives of the represented function.

\textbf{Optimized boundary handling.} Finally, we also propose a boundary handing technique that relies on a classical shift of the solution, but defined by the gradient descent procedure on par with variational angles optimization. This removes the need to include boundary information in the cost expectation, but information still needs to be included in the loss function. %whether via a boundary loss term or via regularization. 
Namely, we seek for the solution in the form
\begin{align}
\label{eq:}
    f(x) = f_c + \langle f_{\varphi,\bm{\theta}}(x) | \hat{\mathcal{C}} |f_{\varphi,\bm{\theta}}(x)\rangle,
\end{align}
where $f_c \in \mathbb{R}$ is a variational parameter alongside the quantum ansatz angles and updated accordingly via the classical optimizer. Therefore, the gradients for $f_c$ have to be calculated additionally when using this boundary handler. One strength of the described method is that, due to the classical shift, even if the random initial angles start such that $\langle f_{\varphi,\bm{\theta}_{\mathrm{init}}}(x_0) | \hat{\mathcal{C}} |f_{\varphi,\bm{\theta}_{\mathrm{init}}}(x_0)\rangle$ is far from the initial value $u_0$, the optimizer can quickly and easily update $f_c$ to rectify this. A weakness however is that the boundary and differential terms in the loss may compete against one another.

\subsection{Regularization}

Given that our goal is to find a variational spectral representation of the differential equations solution using large basis sets, the optimization procedure benefits from having a good initial guess, or ``pre-trained'' DQCs. We can achieve this by introducing a regularization procedure \cite{Neumaier1998,Girosi1995}, also helping the optimizer to avoid getting trapped in local minima.
Variants of the regularization procedure include: 1) feeding-in prior information about the potential solution; 2) biasing the DQC-based solution into a specific form; 3) searching for a solution in a region close to the boundary values, and feeding-in points from the first training into next sessions. The input for procedures 1) and 2) consists of regularization points for the variable(s) $\{ x_{\mathrm{reg}} \}_{r=1}^{R}$, together with corresponding function values $\{ u_{\mathrm{reg}} \}_{r=1}^{R}$ for $R$ points. Similarly, we can consider regularization based on the derivative values. We employ the simplest strategy where an additional contribution to the loss function comes from the regularization points, $\mathcal{L}_{\bm{\theta}}^{\mathrm{(reg)}}[f, x]$. This loss is defined such that the DQC-based function matches the regularisation values at corresponding grid points. This has a form analogous to the boundary loss contribution. Using MSE loss as example, the regularization contribution reads
\begin{align}
\label{eq:reg_loss}
    \mathcal{L}_{\bm{\theta}}^{\mathrm{(reg)}}[f, x] = \sum_{r=1}^{R} \zeta(n_j) \Big(f_{\bm{\theta}}(x_{\mathrm{reg}, r}) - u_{\mathrm{reg}, r} \Big)^2,
\end{align}
where $n_j$ denotes the iteration step. $\zeta(n_j)$ is introduced as an iteration step-dependent regularization weight, and thus denoting an \emph{optimization schedule}. In general, we require higher emphasis on the regularization-based training at initial stages, which shall diminish to zero at higher iteration numbers. This leads to the prior information being used at first, setting a rough solution or preferred function behavior, followed by precise derivative loss optimization at later training stages. 
One possible choice of an optimization schedule corresponds to linearly decreasing regularization weight, $\zeta(n_j) = 1 - n_j/n_{\mathrm{iter}}$,
%
%\begin{align}
%    \zeta(n_j) = 1 - \frac{n_j}{n_{\mathrm{max}}}
%\end{align}
%
where $n_j$ is current iteration number and $n_{\mathrm{iter}}$ is the maximum iteration number. This strategy works for small learning rates and large number of iterations, such that the optimizer has sufficient ``time'' to adjust to the constantly changing loss landscape. Another choice corresponds to a \emph{reverse sigmoid optimization schedule}, where a smooth drop of regularization weight is performed at pre-defined training stages. We parametrize this schedule as
\begin{align}
\label{eq:opt-schedule}
    \zeta(n_j) = 1 - \tanh \left(\frac{n_j - n_{\mathrm{drop}}}{\delta_j n_{\mathrm{iter}}} \right)
\end{align}
where $n_{\mathrm{drop}}$ denotes the iteration step number at which regularisation weight drops, and $\delta_j$ assigns the transition rate. This allows the DQC to initially focus almost entirely on the regularisation optimization, later switching the focus towards the gradient optimization. 
%%%

%%%============================

\subsection{Multifunction encoding}

When solving a \emph{system of} differential equations we need to decide how multiple functions should be encoded simultaneously. There are several ways that this can be achieved. The first approach we consider is to use the same quantum register for all functions, thus compressing information about the function vector using the same feature map circuit and variational ansatz. The functions are then defined through the choice of different cost operators at the readout stage. This method is resource-frugal, and is suitable for certain systems. However, the choice of suitable cost operators becomes complicated, as in some cases shared register encoding may exhibit competition between the optimization of different functions. This concerns the question of expressibility of the set of cost operators, and may potentially be solved using a weighted sum of operators with optimized weights. 
%A good choice of cost functions may be able to mitigate this issue.
A second option is to use separate quantum registers, and correspondingly a different feature map and variational ansatz for each function. This removes the issue of choosing the cost operators, and avoids the related direct competition due to independent parametrization. While requiring more resources we note that function and derivative evaluation can be done in parallel. However, we note that as a combined loss function is considered care must be taken when timing the circuit runs.

\subsection{DQC-based solver: the workflow}

Finally, using the elements and strategies described above, we present a workflow for constructing the differential equation solver based on derivative quantum circuits. This is summarized in Fig.~\ref{fig:workflow} showing the flowchart. We start by specifying the input for the solver. This comprises the problem in hand, specified as a set of nonlinear differential equations of various types, together with their respective boundary conditions. Additionally, a set of regularization points may be added to ensure the optimized solution is chosen in the desired qualitative form. Next, we set up the schedule for derivative quantum circuit optimization and choose the quantum circuit composition. For this we choose: a) the type of quantum feature map; b) the ansatz of variational quantum circuit, including its depth; c) the cost function type, also choosing if variational weights are considered; d) the type of the loss function; e) the strategy to match the boundary terms and derivatives. We also need to specify the classical optimizer for variational angles and weights (with associated hyperparameters), including the number of iterations and exit conditions. Finally, we specify whether the loss function uses a specific optimization schedule, where it is changing during the training.

Once the DQC structure and optimization schedule are defined, we need to specify a set of points $\mathbb{X}$ for each equation variable. This can be a regular equidistant grid, Chebyshev grid, or a randomly-drawn grid. The variational parameters are set to initial values $\bm{\theta} \leftarrow \bm{\theta}_{\mathrm{init}}$ (e.g. as random angles). The expectation value over variational quantum state $|u_{\varphi,\bm{\theta}}(x_i)\rangle$ for the cost function is estimated using the quantum hardware for the chosen point $x_i$. Then a potential solution at this point is constructed, accounting for the boundary handling procedure. Next, the derivative quantum circuits are constructed and their expectation value is estimated for the specified cost function, at point $x_i$. Repeating the procedure for all $x_i$ in $\mathbb{X}$ we collect function values and derivatives, and compose the loss function for the entire grid and system of equations (forming required polynomials and cross-terms by classical post-processing). Regularization points may be also added, biasing the solution to take specific values at these points. The goal of the loss function is to assign a ``score'' to how well the potential solution (parametrized by the variational angles $\bm{\theta}$) satisfies the differential equation, matching derivative terms and the function polynomial to minimize the loss. With the aim to increase the score (and decrease the loss function), we compute the gradient of the loss function with respect to variational parameters $\bm{\theta}$. Using the gradient descent procedure (or in principle any other classical optimization procedure) we update the variational angles from iteration $n_j = 1$ into the next one $n_j + 1$, $\bm{\theta}^{(n_j + 1)} \leftarrow \bm{\theta}^{(n_j)} - \alpha \nabla_{\bm{\theta}} \mathcal{L}$ (with $\alpha$ being here a ``learning'' rate), and repeat the steps outlined before until we reach the exit condition. The exit condition may be chosen as: 1) the maximal number of iterations $n_{\mathrm{iter}}$ reached; 2) loss function value is smaller than pre-specified value; and 3) loss gradient is smaller than a certain value. Once we exit the classical loop, the solution is chosen as a circuit with angles $\bm{\theta}_{\mathrm{opt}}$ that minimize the loss. Finally, we extract the full solution by sampling the cost function for optimal angles $\langle u_{\varphi,\bm{\theta}}(x) | \hat{\mathcal{C}} |u_{\varphi,\bm{\theta}}(x)\rangle$. Notably, this can be done for any point $x$, as DQC constructs the solution valid also beyond (and between) the points at which loss is evaluated originally.
%%%
\begin{figure*}
\centering
\includegraphics[width=1.0\linewidth]{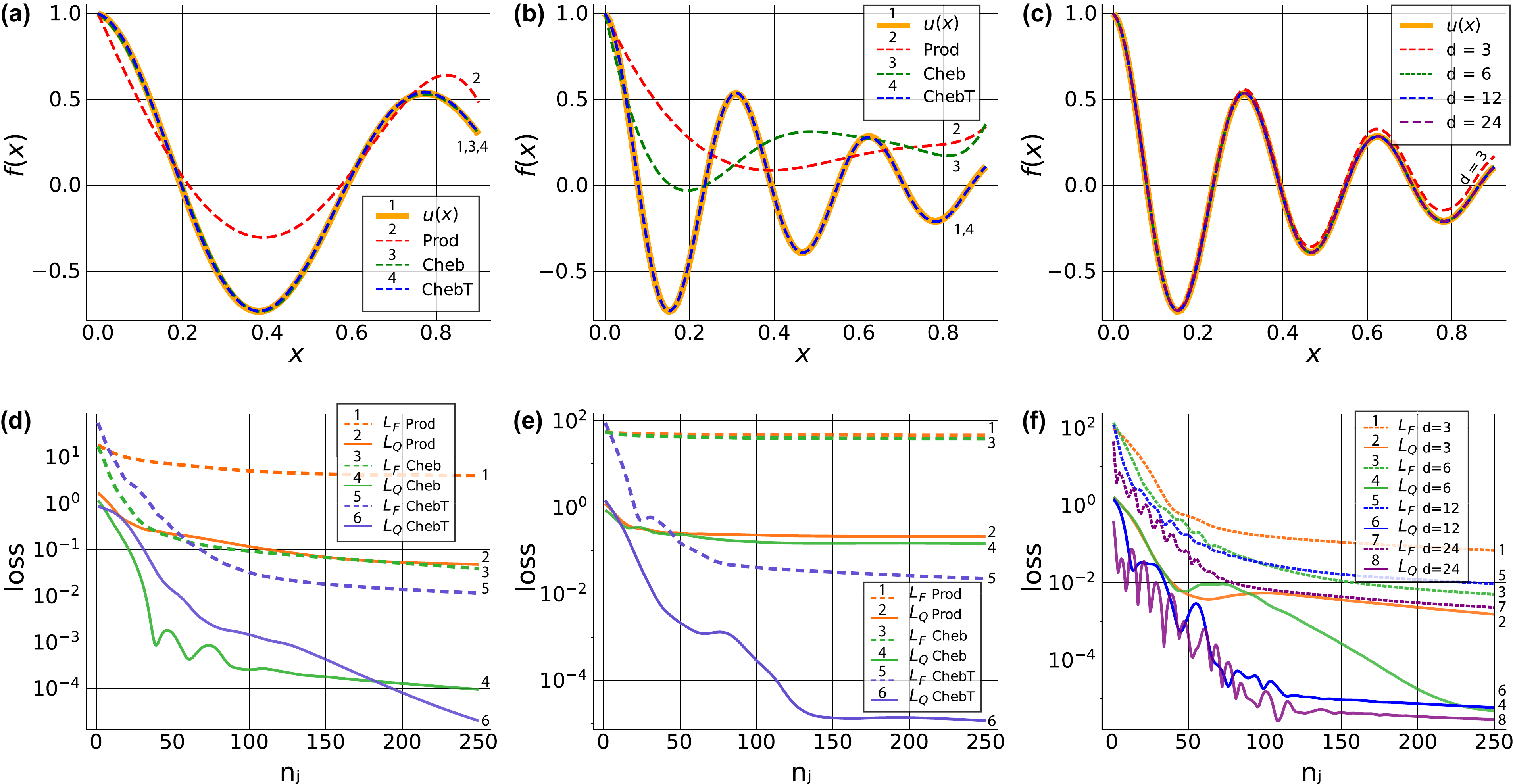}
\caption{\textbf{DQC-based solution for the single ODE example.} (a) Results from the circuit trained to solve Eq.~\eqref{eq:osc_damp} for $u_0 = 1$, $\lambda = 8$, and $\kappa=0.1$. We present DQC solutions $f_{\varphi,\bm{\theta}}(x)$ obtained using three different quantum feature maps $\varphi(x)$ labelled as {\sffamily{}Prod}, {\sffamily{}Cheb}, and {\sffamily{}ChebT}. Corresponding DQC solutions are shown by dashed curves and labelled for clarity (see legend). Analytical solution $u(x)$ is shown by the thick solid curve (label 1). (b) Same as in (a), but for $\lambda = 20$ example. (d, e) \emph{Full loss} $L_{\mathrm{F}}$ (dashed curves) and \emph{quality of solution} $L_{\mathrm{Q}}$ (solid curves) are shown as functions of iteration number $n_j$ for optimization results displayed in (a) and (b), respectively. (c) DQC-based solution of Eq.~\eqref{eq:osc_damp} shown for four different ansatz depths $d = 3, 6, 12, 24$, with the analytical solution $u(x)$ presented by the solid curve. The case of $d=3$ is labelled for clarity. (f) Full loss (dotted curves) and quality of solution (solid curves) shown as a function of iteration number for the solution in (c).}
\label{fig:damp_osc}
\end{figure*}
%%%

%================

\section{Results}

\textbf{Differential equation example.} Now let us see how the algorithm performs in practice. For this we choose a differential equation with a known analytical solution, and compare it to the one obtained by the derivative quantum circuit. We choose a single ODE for the initial value problem which reads
\begin{equation}
\label{eq:osc_damp}
 \frac{du}{dx} + \lambda u \Big(\kappa + \text{tan}(\lambda x) \Big) = 0, ~~~ u(0) = u_0,
\end{equation} 
where $\lambda$ and $\kappa$ are real parameters, and $u_0$ sets the value of the function $u$ at $x=0$. Eq.~\eqref{eq:osc_damp} has a solution in the form of a damped oscillating function, 
\begin{equation}
\label{eq:damp_osc_sol}
u(x) = \text{exp}(-\kappa \lambda x) \text{cos}(\lambda x) + \mathit{const},
\end{equation}
where $\mathit{const}$ is determined by the initial condition.
While the problem is fairly simple being a single ODE, reproducing the damped oscillating solution requires a rich basis of fitting functions, that needs to include both oscillatoric and increasing/decreasing functions. As $\lambda$ and $\kappa$ grow the function starts to oscillate and decay even more rapidly, and the solution becomes harder to express.

To show how the proposed method works, we use derivative quantum circuits to solve Eq.~\eqref{eq:osc_damp} using optimization of differentiable quantum feature maps. Specifically, we choose two cases with parameters $\lambda = 8$ and $\lambda = 20$, and fixed $\kappa = 0.1$, $u_0 = 1$. These problem parameters are chosen to make DQC construction challenging, with $\lambda = 20$ being a complex case as the resulting solution is highly nonlinear and oscillatoric. We consider an equidistant optimization grid of $20$ points, starting from $x = 0$, with maximal time of $0.9$ (dimensionless units are used). This is chosen such that the region with diverging derivative of the nonlinear feature function $d\varphi(x)/dx$ is avoided, and we note that $x$ can be rescaled to match required boundaries. To find the solution we use a quantum register with $N=6$ qubits, and the cost function is chosen as total magnetization in the $Z$ direction, $\hat{\mathcal{C}} = \sum_{j=1}^N \hat{Z}_j$. For the variational circuit, we choose the standard hardware-efficient ansatz described in the Methods section, setting the depth to $d=5$. To search for optimal angles $\bm{\theta}_{\mathrm{opt}}$ we perform adaptive stochastic gradient descent using {\sffamily{}Adam} \cite{adam} with automatic differentiation enabled by analytical derivatives. Specifically, we code the workflow using {\sffamily{}Yao.jl} package \cite{Luo2019,Zeng2019,Liu2019} for {\sffamily{}Julia}, which allows fast and efficient implementation. A full quantum state simulator is used in a noiseless setting. In this example we use the \emph{floating boundary} handling.

We search for the circuit-based solution using three different feature maps described in the Methods section, and compare their performance. These correspond to the \emph{product feature map} [Eq.~\eqref{eq:feature_L1}], the sparse version of the \emph{Chebyshev feature map} [Eq.~\eqref{eq:Cheb-feat-sparse}], and the \emph{tower Chebyshev feature map}, as defined in Eq.~\eqref{eq:Cheb-feat-tower}. We label results for these feature maps as {\sffamily{}Prod}, {\sffamily{}Cheb}, and {\sffamily{}ChebT}, respectively. To assess the performance we use several metrics. The first metric is the \emph{full loss} (denoted as $L_F$). It refers to the loss calculated from the differential equations and any boundary or regularisation terms, in this case $\mathcal{L}_{\bm{\theta}}[d_{x}f, f, x] = \mathcal{L}_{\bm{\theta}}^{\mathrm{(diff)}}[d_{x}f, f, x] + \mathcal{L}_{\bm{\theta}}^{\mathrm{(boundary)}}[f, x] + \mathcal{L}_{\bm{\theta}}^{\mathrm{(reg)}}[f, x]$. The second metric corresponds to \emph{differential loss} ($L_D$), being a part of the full loss excluding regularization contribution. Finally, the third metric is the \emph{quality of solution} ($L_Q$). The quality of the solution is the distance of the current DQC-based solution from the known true solution. This is calculated by evaluating the DQC-based solution and true solution at a set of points and using the MSE loss type, being equal to $L_Q = (1/M) \sum_{i=1}^{M}[f(x_i) - u(x_i)]^2$. Quality of solution gives us a useful way to compare how two different training setups perform, especially if they are training to solve the same differential equations.

The results of DQC training are shown in Fig.~\ref{fig:damp_osc}. In the panel Fig.~\ref{fig:damp_osc}(a, b) we show the solutions of Eq.~\eqref{eq:damp_osc_sol} for $\lambda = 8$ and $\lambda = 20$, respectively, where solid curve (with label 1) represents the analytical solution $u(x)$ in Eq.~\eqref{eq:damp_osc_sol}. The dashed curves (labels 2, 3, 4) represent the final DQC-based solutions sampled from the cost function at approximation points ($n_\mathrm{iter} = 250$ is used). In Fig.~\ref{fig:damp_osc}(d,e) we show the relevant training metrics as a function of iteration number, where solid curves denote the \emph{quality of solution} (curves 2, 4, 6) and dashed curves represent the \emph{full loss} (curves 1, 3, 5). 
We observe that for $\lambda = 8$ both Chebyshev feature maps converge closely to the true solution [Fig.~\ref{fig:damp_osc}(a,d)]. The more expressible \emph{Chebyshev tower feature map} (curves 5, 6) takes longer to converge but reaches a solution closer to the true solution. The less powerful \emph{product feature map} fails to converge with the loss quickly plateauing (curves 1, 2).
For $\lambda = 20$ the true solution is more oscillatoric and has stronger damping, making the solution harder to represent [Fig.~\ref{fig:damp_osc}(b,e)]. The product feature map still fails to converge (curves 1, 2), but now also failing to converge is the simpler Chebyshev feature map (curves 3, 4). The full loss for both cases plateaus rapidly. The more expressible {\sffamily{}ChebT} feature map continues to perform well (curves 5, 6). This supports the hypothesis that choosing a feature map expressible enough for the problem is important, and more simulations with more qubits offers a way to increase the power drastically.
%%%
\begin{figure}
\centering
\includegraphics[width=1.0\linewidth]{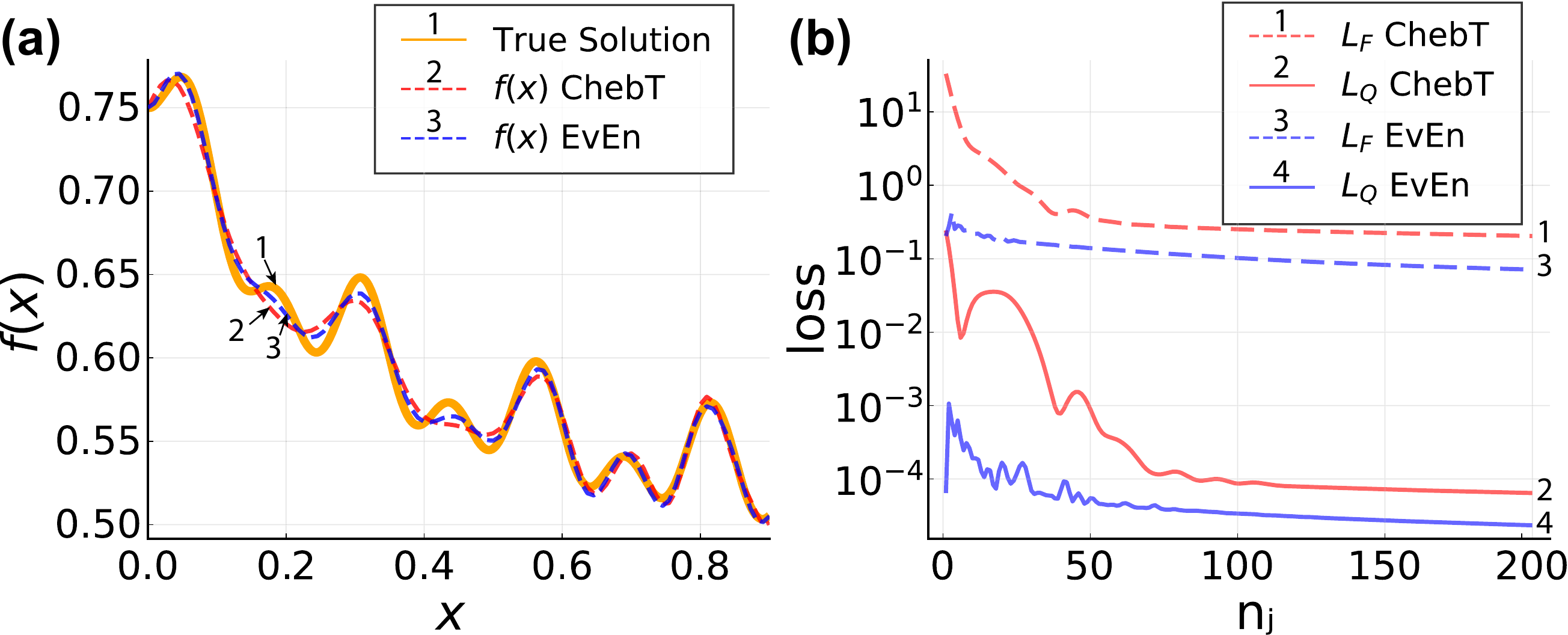}
\caption{\textbf{DQC-based solution for highly nontrivial dynamics example.} (a) Results from the circuit trained to solve initial value problem in Eq.~\eqref{eq:nont_dyn_eq} with $u_0 = 0.75$. We show solutions obtained using two different feature maps: Chebyshev tower feature map ({\sffamily{}ChebT}) and \emph{evolution-enhanced} Chebyshev tower feature map ({\sffamily{}EvEn}). Corresponding DQC solutions are shown by dashed curves (labelled by 2 and 3, respectively). The true solution is shown by a solid curve (labelled by 1). (b) Full loss $L_F$ (dashed curves 1 and 3) and quality of solution $L_Q$ (solid curves 2 and 4) are shown as functions of iteration number $n_j$ for optimization results in (a).}
\label{fig:EEFM}
\end{figure}
%%%

Next, we compare the effect of ansatz depths for the variational circuit $\hat{\mathcal{U}}_{\bm{\theta}}$. We use $d = 3,6,12,24$, $\lambda = 20$ and the \emph{Chebyshev tower feature map}, and the rest of the training setup remains the same as previously considered. These results are presented in Fig.~\ref{fig:damp_osc}(c,f). We observe that for lower depths the solver is slower to converge and does not reach as high accuracy as it does for higher depths. As depth increases more layers of parametrized gates are included in the variational ansatz and so the number of variational angle parameters increase. This causes an increase in the number of gate operations needed in each iteration and how many parameters the classical optimizer needs to update, raising the time taken per iteration. As the depth of the ansatz continues to increase eventually the problem of barren plateaus could be encountered \cite{McClean2018}. Then vanishing gradients would cause the solver to struggle to improve the parameters, however at $d=24$ we had not yet ran into this with over 400 variational parameters. We also note that the alternating blocks ansatz is designed in the way that vanishing gradients can be avoided for certain conditions \cite{Yamamoto2020}.

%%%%%%%%%%%%%%%%%%%%%

\textbf{Differential equation with highly nontrivial dynamics.} To highlight the importance of powerful feature maps for solving differential equations, we provide another example with a rapidly oscillation non-periodic solution. We consider an initial value problem
\begin{equation}
\label{eq:nont_dyn_eq}
\frac{du}{dx} - 4u + 6u^2 - \text{sin}(50x) - u \text{cos}(25x) + 1/2 = 0, ~~~ u(0) = u_0,
\end{equation}
for the function $u(x)$. This differential equation has a solution that is hard to represent since it involves both oscillatoric terms with high frequency and non-oscillatoric terms representing growth and decay. Therefore a feature map with high expressibility is vital for this problem.
%%%
\begin{figure*}
\includegraphics[width=1.0\linewidth]{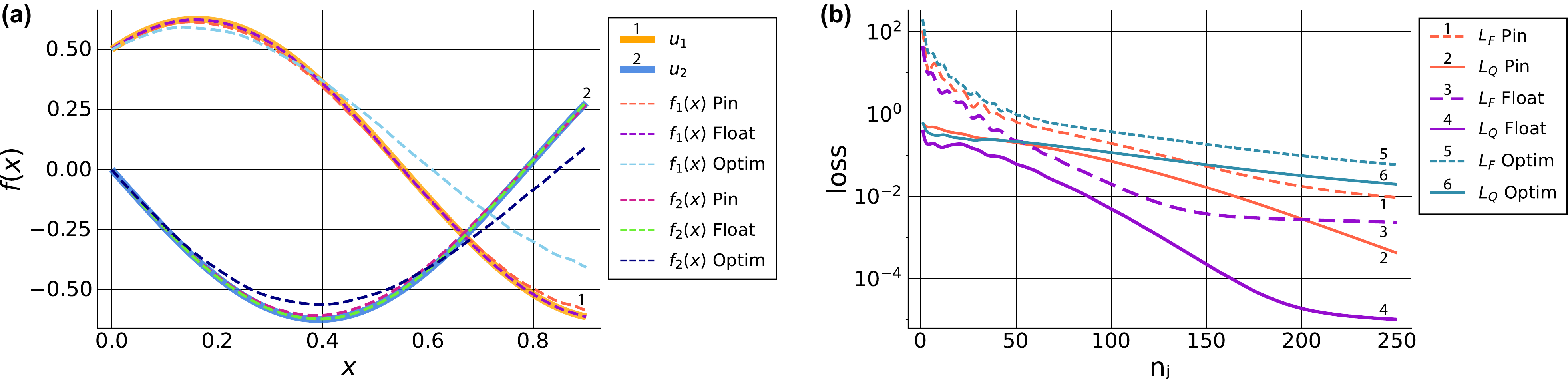}
\caption{\textbf{DQC-based solution for the strongly coupled equations example.} (a) Results from the circuit trained to solve the system of differential equations Eq.~\eqref{eq:coupled_1} and Eq. \eqref{eq:coupled_2} for $u_{1,0} = 0.5$, $u_{2,0} = 0$, $\lambda_1 = 5$, and $\lambda_2 = 3$. We present DQC solutions $f_{1}(x)$ and $f_{2}(x)$ obtained using three different boundary evaluation techniques: pinned boundary (labelled as {\sffamily{}Pin}), floating boundary ({\sffamily{}Float}), and optimized boundary ({\sffamily{}Optim}). Corresponding DQC solutions are shown by dashed curves. Analytical solutions $u_1(x)$ and $u_2(x)$ are shown by thick solid curves. (b) \emph{Full loss} $L_{\mathrm{F}}$ (dashed curves 1, 3, 5) and \emph{quality of solution} $L_{\mathrm{Q}}$ (solid curves 2, 4, 6) are shown as functions of iteration number $n_j$ for optimization results displayed in (a).}
\label{fig:coupled_des}
\end{figure*}
%%%
To illustrate this point, we solve Eq.~\eqref{eq:nont_dyn_eq} using DQCs with both the regular Chebyshev tower feature and an \emph{evolution-enhanced} version of that feature map. As shown in Fig.~\ref{fig:damp_osc}, the Chebyshev tower feature map was the most expressible of the three considered for that example. In this example the evolution-enhanced feature map we consider consists of an application of the Chebyshev tower feature map followed by the unitary evolution $\exp(-i\hat{H}\tau )$ with Hamiltonian $\hat{H}$. Specifically, we have chosen a nearest-neighbor Ising Hamiltonian $\hat{H} = -J \sum_{j} \hat{Z}_j\hat{Z}_{j+1} + h \sum_{j} \hat{X}_j$. The evolution time $\tau$ is fixed, and we consider couplings $J$ and magnetic field $h$ to be drawn uniformly from $(0,1]$. Note that the type of Hamiltonian is not crucial, and may depend on hardware implementation.

For the training regime we consider $u_0 = 0.75$, $100$ training points, and set evolution time $\tau = 2$. We check that results remain consistent for different $J$ and $h$ generated randomly. We use Adam optimizer with learning rate $0.01$ and $200$ iterations. The rest of the training setup is the same as considered for problem Eq.~\eqref{eq:osc_damp} in the \textbf{Differential equation example} with a six-qubit register, cost choice of total magnetization in the $Z$ direction, and a hardware efficient variational ansatz with depth $d = 5$.

The results are shown in Fig.~\ref{fig:EEFM}. First, the circuit is trained with the Chebyshev tower feature map. Being expressive in itself, it manages to come close to the true solution but misses some details for getting a precise solution [Fig.~\ref{fig:EEFM}(a), red dashed curve labelled as 2]. The angles this training results in are then passed as initial angles for training with the evolution-enhanced feature map. The circuit is now further trained for another $200$ iterations to improve upon the solution obtained from the initial training. As can be seen, after this second stage of training, we are able to capture the intricate details of the solution [Fig.~\ref{fig:EEFM}(a), blue dashed curve labelled by 3], and both the full loss and quality of solution are improved upon [Fig.~\ref{fig:EEFM}(b)]. The expressive power of the evolution-enhanced map can be increased for more complex $\hat{H}$ and larger $\tau$, though at the expense of trainability. The full discussion of the evolution-enhanced feature map will be the subject of separate work. 

We note that by adding a quantum evolution we increase expressibility of the circuit. This helps to represent highly oscillatoric functions. At the same time, DQCs in this case are more difficult to train, as keeping track of the gradients for oscillatoric function requires a finer training grid and has a slower convergence. We envisage that $\tau[n_j]$ can be set as a \emph{flowing} parameter of the feature map that increases from zero as iteration number $n_j$ grows (similar to annealing procedure as in adiabatic quantum computing). This will ensure easy trainability at the start, followed by adjusting $\bm{\theta}$'s for a larger fitting function set providing extra precision. Another approach is to take $\tau(x)$, adding variable-dependent amplitudes directly through the evolution. While in this case parameter shift rule is not applicable, the circuit differentiation can be done numerically \cite{Bespalova2020}, with wavefunction overlaps measured with the SWAP test \cite{MitaraiPRRes}.

%%%%%%%%%%%%%%%%%%%%%

\textbf{Strongly coupled equations.} Building up on the single ODE example, we proceed to consider a \emph{system} of differential equations, taking two strongly coupled differential equations as an example. This describes the evolution of competing modes $u_1(x)$ and $u_2(x)$ as a function of variable $x$, which in this case corresponds to time. The associated rate equations read
\begin{align}
    \label{eq:coupled_1}
    F_1[d_x \bm{u}, \bm{u}, x] = \frac{du_1}{dx} & - \lambda_1 u_2 - \lambda_2 u_1 = 0,~~ u_1(0) = u_{1,0},\\
    F_2[d_x \bm{u}, \bm{u}, x] = \frac{du_2}{dx} & + \lambda_2 u_2 + \lambda_1 u_1 = 0,~~ u_2(0) = u_{2,0},
    \label{eq:coupled_2}
\end{align}
where $\lambda_{1,2}$ are coupling parameters, $u_{1,0}$, $u_{2,0}$ are initial conditions. The larger $|\lambda_1|$ is in comparison to $|\lambda_2|$ the more strongly coupled the two equations will be. This can be intuitively seen considering $|\lambda_1| \geq |\lambda_2|$, leading to larger contribution of $u_2$ into the equation for $du_{1}/dx$ derivative than $u_1$, and vice versa.
%%%
\begin{figure*}
\centering
\includegraphics[width=1.0\linewidth]{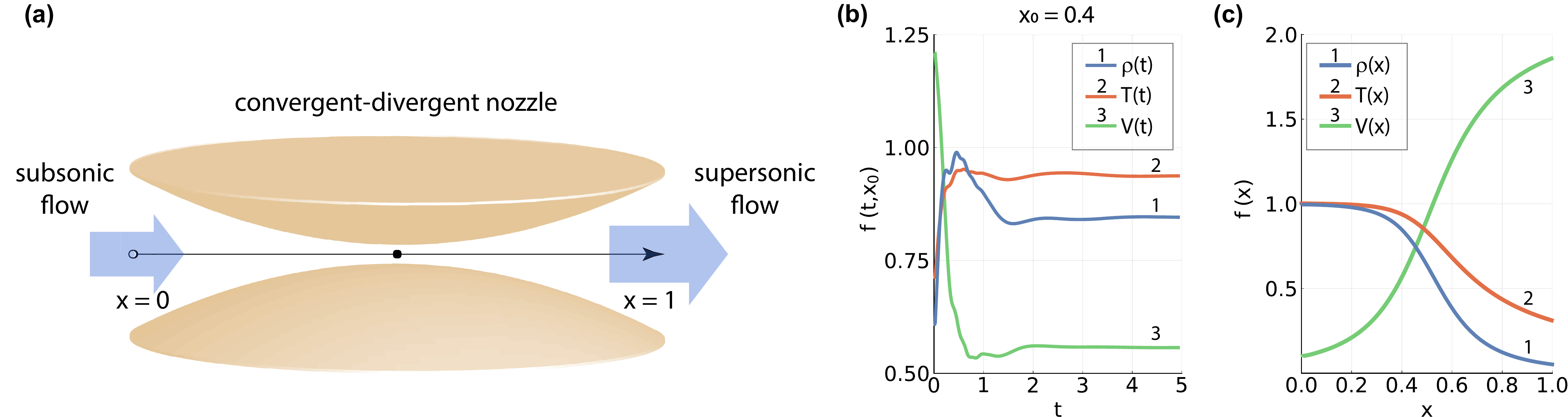}
\caption{\textbf{Quasi-1d fluid dynamics.} (a) We consider an example of fluid dynamics problem corresponding to a convergent-divergent nozzle. The air flows from converging part of the nozzle ($x < 0.5$), passes through the throat placed symmetrically in the middle, and exits to diverging part ($x > 0.5$). (b) System variables [density $\rho(t)$, temperature $T(t)$, and velocity $V(t)$] are shown as functions of time, near the centre of the nozzle at $x_0=0.4$. For clarity, we label the corresponding curves by 1, 2, and 3. (c) Steady state solutions are plotted as functions of the spatial dimension $x$, with the same labelling.}
\label{fig:nozzle}
\end{figure*}
%%%

To move from considering one differential equation to a system of equations we need to encode multiple functions using quantum registers as described in the Sec.~III~Methods. For this specific example we choose a simple parallel encoding, where separate cost functions and ansatze are considered. In this case each function has a separate set of parameters to be optimized. The loss is changed accordingly to include information on separate contributions from two coupled differential equation that are optimized simultaneously.
We encode each function using the differentiable feature map combined with individual variational ansatz parametrized by the set of angles $\bm{\theta}_{1}$ and $\bm{\theta}_2$ and before deciding on the boundary evaluation type we have
\begin{align}
&f_1(x) = \langle \text{\O} | \hat{\mathcal{U}}_{\phi}^\dagger (x) \hat{\mathcal{U}}_{\bm{\theta}_{1}}^\dagger \hat{\mathcal{C}}^{(1)} \hat{\mathcal{U}}_{\bm{\theta}_{1}} \hat{\mathcal{U}}_{\phi}(x) | \text{\O} \rangle,\\
&f_2(x) = \langle \text{\O} | \hat{\mathcal{U}}_{\phi}^\dagger (x) \hat{\mathcal{U}}_{\bm{\theta}_{2}}^\dagger \hat{\mathcal{C}}^{(2)} \hat{\mathcal{U}}_{\bm{\theta}_{2}} \hat{\mathcal{U}}_{\phi} (x) | \text{\O} \rangle,
\end{align}
%
% multi-cost version
%\begin{align}
%&f_1(x) = \langle \text{\O} | \hat{\mathcal{U}}_{\phi}^\dagger (x) \hat{\mathcal{U}}_{\bm{\theta}_{1}}^\dagger \Big( \sum_\ell^L \alpha_\ell^{(1)} \hat{\mathcal{C}}_i^{(1)} \Big) \hat{\mathcal{U}}_{\bm{\theta}_{1}} \hat{\mathcal{U}}_{\phi}(x) | \text{\O} \rangle,\\
%&f_2(x) = \langle \text{\O} | \hat{\mathcal{U}}_{\phi}^\dagger (x) \hat{\mathcal{U}}_{\bm{\theta}_{2}}^\dagger \Big( \sum_\ell^L \alpha_\ell^{(2)} \hat{\mathcal{C}}_i^{(2)} \Big) \hat{\mathcal{U}}_{\bm{\theta}_{2}} \hat{\mathcal{U}}_{\phi} (x) | \text{\O} \rangle.
%\end{align}
%Additionally we introduced multi-term cost functions such that gradient descent can separately adjust coefficients $\alpha_\ell^{(u_1,u_2)}$ for the two functions. 
%
where $\hat{\mathcal{C}}^{(1,2)}$ are in principle different cost functions for each equation. 
For the loss function we consider the sum of the MSE losses for the first and second differential equation. This loss is written as
$\mathcal{L}_{\bm{\theta}}[d_{x}\bm{f}, \bm{f}, x] = \sum_{i=1}^M L\big( F_1[d_x \bm{f}, \bm{f}, x_i], 0 \big) + \sum_{i=1}^M L\big( F_2[d_x \bm{f}, \bm{f}, x_i], 0 \big)$ with $F_1$ and $F_2$ as written in Eq.~\eqref{eq:coupled_1}-\eqref{eq:coupled_2}, and $L$ depends on the loss choice as detailed in the Methods section. There can be additional boundary loss terms depending on boundary evaluation method chosen. If present, they contribute to the loss function as a sum of the boundary terms for $u_1$ and $u_2$. Note that we consider the loss as a sum of individual contributions coupled together we are trying to minimise both simultaneously with equal weight. Due to the coupling between the two equations this could lead to competition between the two loss terms (a parameter update which causes a loss decrease for one DE's loss may lead to an increase in the other). This may result in increasing the chance of converging to a local minima rather than the global minimum; however, this effect can be mitigated if loss contributions are weighted in some way. Another solution is to use quantum kernel methods, where loss corresponds to the overlap between quantum feature states. 
%for example one DEs path to it's solution in the global minima may require the loss term of the other DE to briefly increase causing the optimizer to avoid that route. 
Choosing a suitable loss function in this case is an important point to consider in the future.

We define the problem setting parameters to $\lambda_1 = 5, ~ \lambda_2 = 3$ and initial conditions to $u_{1,0} = 0.5, ~ u_{2,0} = 0$. We set up the training scheme as in \textbf{Differential equation example} subsection using a six-qubit register, cost choice of total magnetization in the $Z$ direction for both $u_1$ and $u_2$, hardware-efficient variational ansatz with depth $d=5$, {\sffamily{}Adam} optimizer with learning rate $0.02$, and feature map choice of the \emph{Chebyshev tower} feature map. We test the performance for the three boundary evaluation types: \emph{pinned boundary}, \emph{floating boundary}, and \emph{optimized boundary}.
The results are shown in Fig.~\ref{fig:coupled_des}. The pinned and optimized boundary handlers perform similarly, slowly converging to the analytical solution $u_{1,2}(x)$ within $250$ iterations [Fig.~\ref{fig:coupled_des}(a)].  The two approaches have similar convergence in terms of the \emph{full loss} [Fig.~\ref{fig:coupled_des}(b), dashed curves labelled by 1 and 5], but differ in terms of \emph{quality of solution} [Fig.~\ref{fig:coupled_des}(b), dashed curves labelled by 2 and 6]. When using floating boundary type a function close to the true solution is obtained [with $L_Q$ value of approximately $10^{-5}$, see curve 4 in Fig.~\ref{fig:coupled_des}(b)]. This difference in convergence rate is a result of boundary information having an impact on the loss, demanding the matching for pinned and optimized boundary handlers, whereas the floating boundary automatically matches the initial condition and no loss boundary term is needed. The consequence of competing terms in the loss function can be seen in the early oscillations of the \emph{full loss} in Fig.~\ref{fig:coupled_des}(b). \\

%%%%%%%%%%%%%%%%%%%%%

\textbf{Fluid dynamics applications.} An area where solvers for complex differential equations are much required is fluid dynamics \cite{AndersonBook}. In this case several outstanding models are hard to tackle due to their nonlinear nature. Examples include Burger's equation and Navier-Stokes equations. We concentrate on the latter and show how one can approach them with the DQC solver.
%%%
\begin{figure*}
\centering
\includegraphics[width=1.\linewidth]{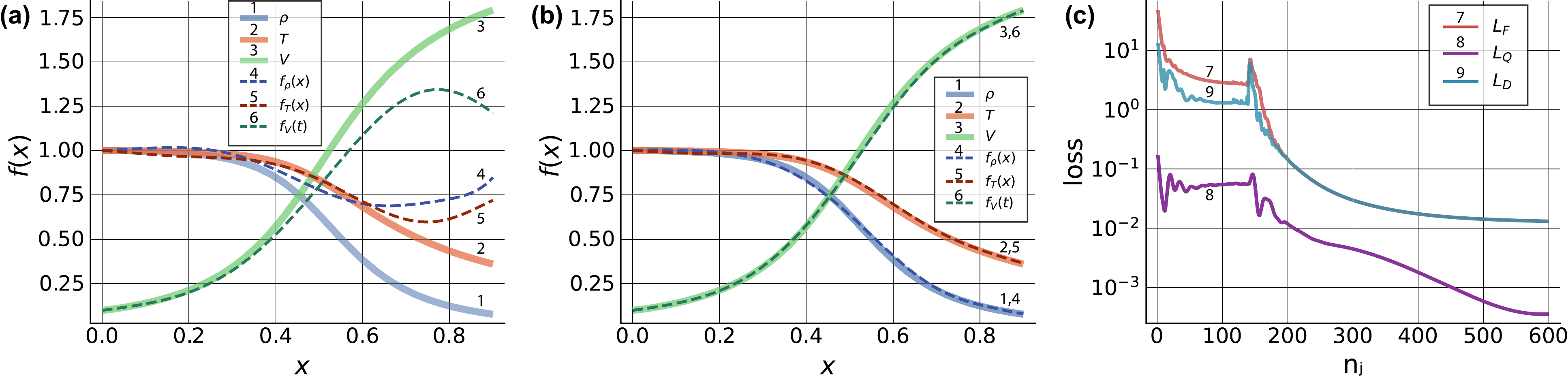}
\caption{\textbf{DQC-based solution for the Navier-Stokes convergent-divergent nozzle problem.} (a) Intermediate solution in training where dashed curves correspond to solutions drawn from trained DQC and solid curves to the true solution. Circuits are trained over twenty points in range $(0, 0.4)$ (b) Final solution. Solid curves with labels 1, 2, and 3 show the true solutions for the density, temperature, and velocity, respectively. Dashed curves with labels 4, 5, and 6 show the DQC solutions for the density, temperature, and velocity, respectively. The DQC solution matches the known solution and also what would be physically expected. As the air goes through the nozzle it accelerates and cools down. (c) Full loss ($L_F$, curve 7), quality of solution ($L_Q$, curve 8), and differential loss ($L_D$, curve 9) are shown as functions of the iteration number $n_j$ for the training resulting in solution (b).}
\label{fig:NS-QNN}
\end{figure*}
Navier-Stokes equations describe a flow of incompressible fluids. This highly nonlinear set of partial differential equations is used to model fluids, magnetoplasma, turbulence etc. It is heavily used in the aerospace industry and weather forecasting.  It can be derived from general principles. Namely, we consider fluid motion that obeys Newton's law and we simply track the fluid mass passing through a (infinitesimal) volume.

The general form Navier-Stokes equation can be presented in the form
\begin{align}
    \frac{\partial (\rho \varv_x) }{\partial t} + \nabla \cdot (\rho \varv_x \mathbf{V}) &= -\frac{\partial p} {\partial x} + \frac{\partial \tau_{xx}} {\partial x} + \frac{\partial \tau_{yx}} {\partial y} + \frac{\partial \tau_{zx}} {\partial z} + \rho f_x ,\\
    \frac{\partial (\rho \varv_y) }{\partial t} + \nabla \cdot (\rho \varv_y \mathbf{V}) &= -\frac{\partial p} {\partial y} + \frac{\partial \tau_{xy}} {\partial x} + \frac{\partial \tau_{yy}} {\partial y} + \frac{\partial \tau_{zy}} {\partial z} + \rho f_y ,\\
    \frac{\partial (\rho \varv_z) }{\partial t} + \nabla \cdot (\rho \varv_z \mathbf{V}) &= -\frac{\partial p} {\partial z} + \frac{\partial \tau_{xz}} {\partial x} + \frac{\partial \tau_{yz}} {\partial y} + \frac{\partial \tau_{zz}} {\partial z} + \rho f_z ,
\end{align}
where $(\varv_x, \varv_y, \varv_z)$ are instantaneous velocities in the $(x,y,z)$ directions, $\tau$ is the stress tensor, $f$ the body force per unit mass acting on the fluid element, $\rho$ the density, and $p$ the pressure. Here $\mathbf{V}$ is a velocity field.

The building blocks for the Navier-Stokes equations described above are the continuity equation for the density $\rho$ subjected to energy conservation and the momentum conservation rules. Finally, as we use energy conservation, we can rewrite equations in terms of one of the thermodynamic state functions, being temperature, or pressure, or enthalpy etc.

While general by itself, Navier-Stokes equations are usually solved in relevant limiting cases. Specifically, this can correspond to space reduction (2D, quasi-1D, 1D), isotropic/anisotropic media properties, and fluid properties (viscous or inviscid flow). The specific example we choose to start with is the flow through a convergent-divergent nozzle, being a paradigmatic task in the aerospace industry [Fig.~\ref{fig:nozzle}(a)] \cite{Gaitan2020,AndersonBook}. The Navier-Stokes equations can be rewritten for the inviscid fluid in quasi-1D approximation. They read \cite{AndersonBook,Gaitan2020}
\begin{align}
\label{eq:continuity}
&\frac{\partial \rho}{\partial t} = -\rho \frac{\partial V}{\partial x} - \rho V \frac{\partial (\ln A)}{\partial x} - V \frac{\partial \rho}{\partial x},\\
\label{eq:energy}
&\frac{\partial T}{\partial t} = -V \frac{\partial T}{\partial x} - (\gamma -1) T \left( \frac{\partial V}{\partial x} + V \frac{\partial (\ln A)}{\partial x} \right),\\
\label{eq:momentum}
&\frac{\partial V}{\partial t} = -V \frac{\partial V}{\partial x} - \frac{1}{\gamma} \left( \frac{\partial T}{\partial x} + \frac{T}{\rho} \frac{\partial \rho}{\partial x} \right),
\end{align}
where Eq.~\eqref{eq:continuity} corresponds to the continuity equation, Eq.~\eqref{eq:energy} describes the energy conservation, and Eq.~\eqref{eq:momentum} stems from momentum conservation. $A(x)$ corresponds to the spatial shape of the nozzle, and is a function of the lateral coordinate $x$. $\gamma$ describes the ratio of specific heat capacities, and is equal to $1.4$ for the relevant case of air flow. Here we used nondimensional variables \cite{AndersonBook}.

We set the problem with nozzle shape
\begin{equation}
A(x) = 1 + 4.95 (2 x - 1)^2,~~ 0\leq x \leq 1,
\end{equation}
and for simplicity specify boundary conditions as
\begin{equation}
\rho(x=0) = 1, \quad T(x=0) = 1, \quad V(x=0) = 0.1,
\end{equation}
when solving the initial value problem for the steady-state flow. We also need to specify the initial conditions if solving the dynamical problem. These are chosen as \cite{AndersonBook}
\begin{align}
\label{eq:NS_init_1}
&\rho(x, t=0) = 1 - 0.944 x,\\
\label{eq:NS_init_2}
&T(x, t=0) = 1 - 0.694 x,\\
\label{eq:NS_init_3}
&V(x, t=0) = (0.1 + 3.27 x) T(x, t= 0 )^{1/2}.
\end{align}

Let us consider the stationary state problem represented by Eqs.~\eqref{eq:continuity}-\eqref{eq:momentum} with conditions above, and equate the time derivatives to zero. Interestingly, when trying to solve the system for the steady state solution using various classical methods, as for example implemented in {\sffamily{}Mathematica's NDSolve}, the calculations do not converge. This proves to be challenging as the system is stiff. Solutions may become unstable depending on initial values, and specifically the input velocity. To understand the problem, it is instructive to rewrite the system of stationary Navier-Stokes equations in the form
\begin{align}
\label{eq:continuity_ss}
&\frac{d \rho}{dx} =  \frac{\rho V^2 d_x (\ln A)}{T - V^2},\\
\label{eq:energy_ss}
&\frac{d T}{dx} = \frac{T V^2 (\gamma - 1) d_x(\ln A)}{T - V^2},\\
\label{eq:momentum_ss}
&\frac{d V}{dx} = -\frac{T V d_x(\ln A)}{T - V^2}.
\end{align}
Immediately we observe that each function at the RHS diverges at the point $x$ s.t. $T(x) = V(x)^2$. This leads to singular behaviour and breaks classical solvers, including the ones with stiff handling. 
At the same time, full dynamical solution and its $t\rightarrow \infty$ extrapolation are possible, shown in Fig.~\ref{fig:nozzle}(b,c) once the \emph{well-suited} initial conditions \eqref{eq:NS_init_1}-\eqref{eq:NS_init_3} are chosen allowing to avoid instability, as usually done in computational fluid dynamics \cite{AndersonBook, Gaitan2020}.\\

\textbf{DQC solution.} We proceed to solve the stationary system of Navier-Stokes equations for the convergent-divergent nozzle by constructing optimized DQC. We consider the case of subsonic-supersonic transition, where flow velocity increases after the center of the nozzle and qualitative behaviour of other state variables (temperature and density drop) is known. However, getting the quantitative results is difficult. We show that derivative quantum circuits can find solutions despite the challenge for the classical solution for the continuous grid.

To construct the solution we employ a two-stage optimization approach, where the solution is first obtained at smaller $x$, and later generalized until the end of the nozzle. For the circuit we again consider a six-qubit quantum register with a Chebyshev quantum feature map, keeping in mind that $x \in [0,1)$. We choose the cost functions as a total magnetization $\hat{\mathcal{C}} = \sum_{j=1}^N \hat{Z}_j$. As we consider three functions $\{\rho(x)$, $T(x)$, $V(x)\}$, three equations contribute to the loss function in the combined manner. We use \emph{floating boundary} handling, where each curve is adjusted according to starting values, chosen as $\rho(0)=1$, $T(0)=1$, $V(0)=0.1$. The variational ansatz is taken in the standard hardware efficient form with $d=6$ depth. {\sffamily{}Adam} is used as a classical optimizer, and the learning rate is set to $0.01$.

At the first stage we train DQC in the region $(x_{\mathrm{min}}, x_{\mathrm{max}}) = (0., 0.4)$, such that the circuit represents the region close to initial point $x=0$. As expected, in the subsonic region flow velocity grows towards the middle of the nozzle, while temperature and density drop slowly [see Fig.~\ref{fig:NS-QNN}(a), dashed curves 4, 5, 6]. The training is set for $20$ equally distributed points of $x$, with no prior regularization and using floating boundary. We optimize DQC for $n_{\mathrm{iter}} = 200$ iterations for {\sffamily{}Adam} with the learning rate of $\alpha = 0.01$. We find a high-quality solution based on the gradient information in the imposed solution region $x<0.4$ shown in Fig.~\ref{fig:NS-QNN}(a).

We proceed to search for the full solution that includes the divergent nozzle part for $x > 0.5$ at the second training stage. At this session we choose the grid of $40$ points, where two regions of $(0, 0.4)$ and $(0.6, 0.9)$ with $20$ points each are used. As we discussed before, the key problem of the convergent-divergent nozzle in the subsonic-supersonic transition case is the divergence around the middle of the nozzle. This causes a major problem to classical solvers, that are unable to find a steady state solution directly. Divergent contributions from this region also impact the loss function, and makes training complicated. However, by excluding the region around the nozzle throat, $(0.4, 0.6)$, in the training we alleviate this problem. Proceeding with the use of the same ansatz and boundary handing, we feed the variational angles from stage 1 as initial parameters for stage 2 training. We employ weak regularization to ensure that the required subsonic-supersonic solution is made. Namely, we use $20$ points in the $(0, 0.4)$ region benefiting from the previously found solution (in general we have access to as many points as we need), and also add $5$ points in the $x \in (0.6, 0.9)$ region representing weak bias towards supersonic solution type. The training is performed for $n_{\mathrm{iter}} = 600$, learning rate of $\alpha = 0.005$, and regularization switch-off function $\zeta(n_j)$ set to be removed smoothly around $n_j=150$. The full solution is shown in Fig.~\ref{fig:NS-QNN}(b). It converges to the expected long-time behavior for the system, where function derivatives from DQC match the nonlinear contributions. The increase of speed for the air flow in the convergent part and decrease of temperature and density is reproduced quantitatively. The details of optimization run can be inferred from the loss plotted in Fig.~\ref{fig:NS-QNN}(c). This shows a distinct region in the presence of regularization ($n_{j} < 150$), where quality of solution $L_Q$ remains low [Fig.~\ref{fig:NS-QNN}(c), curve 8], while the full loss $L_F$ improves [Fig.~\ref{fig:NS-QNN}(c), curve 7] thanks to regularization contribution and weakly improved derivative contribution $L_D$ [Fig.~\ref{fig:NS-QNN}(c), curve 9]. For the region with switched off regularization we observe steady improvement of all metrics, showing that DQCs are efficiently trained to approach true solution, also evidenced by $L_Q$ decrease. Notably, as compared to many methods relying on sparse discretization of $x$, we have found the solution along the full nozzle length.

%===============

\section{Summary and Outlook}
We presented a general framework for solving general (systems of) nonlinear differential equations using differentiable quantum circuits on gate-based quantum hardware. 
The method makes use of quantum feature map circuits to encode function values into a latent space. This allows us to consider spectral decompositions of the trial solutions to differential equations.
We showed how our method can accurately represent nonlinear solutions, using the high-dimensional Hilbert space of a qubit register. For this we exploit as an example a large spectral basis set of Chebyshev or Fourier polynomials.   
We also showed how analytical circuit differentiation can be used to represent function derivatives appearing in differential equations, and constructed loss functions which aim to improve the prepared trial solution.
This method opens up a new way of solving general, complex, (non)linear systems; as an example we presented solutions to the Navier-Stokes equations, but the same method can be applied across all disciplines where differential equations arise.
Although we showed how the method works explicitly for a spectral (global) treatment, the scheme is general and one may instead consider a \emph{digital quantum feature map} which allows a grid-based (local) treatment.
We did not yet consider the impact of noise on the algorithm performance; however, it is well-known that the analytical gradients, which are used here, are much less susceptible to noise than numerical gradients evaluated on variational quantum circuits. We expect recent developments in noisy black-box optimization algorithms for NISQ application would perform well in this method, as well as error mitigation techniques.
The presented quantum feature maps are a good start, but could be improved upon further, where the goal is to find efficient representations of functions. This choice will likely be problem-specific and in some cases problem-motivated, allowing to make optimal use of (quantum) resources. 
A good choice of variational ansatz is crucial to loss function convergence success and speed. We presented some examples, but an active area of research is to improve upon these using for example adaptively growing circuits or stronger interplay with intelligent classical optimization protocols, such as Bayesian optimization.

An important question to discuss is the potential of achieving a quantum advantage, specifically for solving differential equations with a hybrid workflow. Similarly to classical ML, variational quantum machine learning uses quantum circuits as universal function approximators \cite{Goto2020}. While VQA inherently remains a heuristic, similarly to machine learning-type approaches, they are highly suitable to problems where traditional methods fail or are inefficient. For instance, classical neural networks based on Fourier representation were recently shown to outperform traditional PDE solvers by three orders of magnitude \cite{ZongyiLi2020}. This can be pushed further by quantum variational protocols. For expressive feature maps, the representation power of quantum circuits grows superpolynomially with the number of qubits. In classification tasks this leads to increased accuracy even for small networks \cite{Chen2020}. For differential equations, rich basis sets allow for accurate solution representation in a memory-frugal form --- a saved list of optimal circuit parameters for a given DQC structure.
Having the expressibility advantage, the success of operation then depends on: 1) trainability of quantum circuits; 2) absolute operation rates. The former is well-known to suffer from a phenomenon commonly referred to as barren plateaus. That can be mitigated by a carefully chosen circuit structure, as recently shown for convolutional quantum circuits \cite{Cong2019,Pesah2020}. The latter requires quantum hardware specifications with high sampling rates and a low-latency hybrid operational mode. In the pre-FTQC era with large-scale quantum devices, yet limited by noise, parallel operation may offer this possibility. 

Finally, differential quantum circuits can enjoy another type of computational advantage related to training. In this paper we considered capturing the function similarity through a loss function based on a cost Hamiltonian suitable for near-term operation. Anticipating the appearance of quantum devices with deep quantum circuits and problems with multi-dimensional functions, we consider possible strategies for parallel DQC training at different nodes of the training grid. One possibility is training of differential equation solutions as a superposition of quantum circuits (feature maps applied for different values of variables). The DE is then solved by matching differentiated version at all points (using linear combinations of unitaries), and loss can be measured from quantum kernel estimation. This was used recently to speed up generative modelling with quantum batch training \cite{Huang2020}. We will further investigate that approach in future work. To conclude, we suggest that the DQC method could be a potential contender for exploring the boundaries of quantum advantage in industrially relevant problems, and expect many improvements can be made building on the presented idea.

\textit{Acknowledgement.} We thank Benno Broer for useful discussions on the subject and reading the manuscript.

%\textit{Ethics declaration.} A patent application for the method described in this manuscript has been submitted by Qu\&Co BV with OK, AP and VE as inventors.

%%%%%%%%%%%%%%%%%%

%---------------------------------


\begin{thebibliography}{99}
\bibitem{patent} A patent application for the method described in this manuscript has been submitted by Qu \& Co BV with OK, AP and VE as inventors.
\end{thebibliography}
\end{document}